\documentclass[12pt]{aastex}

\shorttitle{The 5.25 and 5.7 $\mu$m PAH Emission Features}

\shortauthors{C. Boersma et al.}

\begin{document}


\title{The 5.25 \& 5.7 $\mu$m Astronomical Polycyclic Aromatic Hydrocarbon Emission Features\footnotemark[1]}\footnotetext[1]{Based on observations with ISO, an ESA project with instruments funded by ESA member states (especially the PI countries: France Germany, the Netherlands and the United Kingdom) and with the participation of ISAS and NASA}

\author{C. Boersma\altaffilmark{2}}
\email{C.Boersma@astro.rug.nl}

\author{A.L. Mattioda\altaffilmark{3,4}} \author{C.W. Bauschlicher Jr.\altaffilmark{5}} \author{E. Peeters\altaffilmark{4,6}} \author{A.G.G.M. Tielens\altaffilmark{2,3}} \author{L.J. Allamandola\altaffilmark{3}}

\altaffiltext{2}{Kapteyn Astronomical Institute, University of Groningen, P.O. Box 800, 9700 AV, Groningen, The Netherlands}
\altaffiltext{3}{NASA Ames Research Center, MS 245-3, Moffett Field, CA 94035, USA}
\altaffiltext{4}{SETI Institute, 515 N. Whisman Road, Mountain View, CA 94043, USA}
\altaffiltext{5}{NASA Ames Research Center, MS 230-3, Moffett Field, CA 94035, USA}
\altaffiltext{6}{Department of Physics and Astronomy, PAB 213, The University of Western Ontario, London, ON N6A 3K7, Canada}

\begin{abstract}

  Astronomical mid-IR spectra show two minor PAH features at 5.25 and
  5.7 $\mu$m (1905 and 1754 cm$^{\rm - 1}$) that hitherto have been
  little studied, but contain information about the astronomical PAH
  population that complements that of the major emission bands. Here
  we report a study involving both laboratory and theoretical analysis
  of the fundamentals of PAH spectroscopy that produce features in
  this region and use these to analyze the astronomical spectra. The
  ISO SWS spectra of fifteen objects showing these PAH features were
  considered for this study, however only four (HD 44179; NGC 7027;
  Orion Bar, 2 positions) have sufficient signal-to-noise between 5
  and 6 $\mu$m to allow for an in-depth analysis. All four
  astronomical spectra show similar peak positions and profiles. The
  5.25 $\mu$m feature is peaked and asymmetric, with a FWHM of about
  0.12 $\pm$ 0.01 $\mu$m ($\sim$40 $\pm$ 6.5 cm$^{\rm -1}$), while the
  5.7 $\mu$m feature is broader and flatter, with a FWHM of about 0.17
  $\pm$ 0.02 $\mu$m (50 $\pm$ 5.6 cm$^{\rm -1}$). Detailed analysis of
  the laboratory spectra and quantum chemical calculations show that
  the astronomical 5.25 and 5.7 $\mu$m bands are a blend of
  combination, difference and overtone bands primarily involving CH
  stretching and CH in-plane and CH out-of-plane bending fundamental
  vibrations. The experimental and computational spectra show that, of
  all the hydrogen adjacency classes possible on PAHs, solo and duo
  hydrogens consistently produce prominent bands at the observed
  positions whereas quartet hydrogens do not. In all, this study
  supports the picture that astronomical PAHs are large with compact,
  regular structures. From the coupling with primarily strong CH
  out-of-plane bending modes one might surmise that the 5.25 and 5.7
  $\mu$m bands track the neutral PAH population. However, theory
  suggests the role of charge in these astronomical bands might also
  be important.

\end{abstract}


\section{Introduction}
\label{sec:introduction}

Some thirty years of observations, combined with computational and
laboratory studies, have shown that the mid-IR astronomical emission
features, formerly referred to as the Unidentified Infrared (UIR)
bands, are produced by mixtures of highly vibrationally excited
Polycyclic Aromatic Hydrocarbons (PAHs) and closely related
species. Detected in many Galactic and extra-galactic objects,
including several with significant redshift
\citep[e.g.][]{2005ApJ...628..604Y}, the astronomical infrared
emission features present an important and unique probe of
astrochemical and astrophysical conditions across the universe. Recent
reviews and papers of the observational and laboratory work
\citep[e.g.][]{2004ApJ...617L..65P, 2004ASPC..309..665H,
  2004ARA&A..42..119V, 2007ApJ...659.1338S, 2007ApJ...656..770S,
  2008ARA&A..46..289T} and work on theoretical models
\citep[e.g.][]{2001A&A...372..981V, 2001ApJ...556..501B,
  2001ApJ...554..778L, 2002A&A...388..639P, 2006A&A...460..519R,
  2007ApJ...657..810D} can be found elsewhere.

The major features at 3.3, 6.2, `7.7', 8.6 and the complex of bands
between 11 and 20 $\mu$m have been studied in great detail
\citep[e.g.][and references therein]{2000A&A...357.1013V,
  2001A&A...370.1030H, 2002A&A...390.1089P, 2004ApJ...611..928V}, and
the fundamental spectroscopic information is now available with which
one can analyze the strongest astronomical features. However, there
are several components of the astronomical PAH emission spectra that
have been widely overlooked. Many of these contain valuable, sometimes
subtle, information which is equally important to that revealed by the
more well-known features. This study focuses on just such features,
namely the weak bands that fall between 5 and 6 $\mu$m (2000 and 1667
cm$^{\rm -1}$).

One of the early predictions of the PAH
hypothesis was the expectation of a weak emission feature near 5.25
$\mu$m in all objects showing the major PAH bands. Its detection in
1989 (\citeauthor{1989ApJ...345L..59A}) was an early confirmation of
the PAH hypothesis. Since that time, although evident in many spectra
showing the major PAH features, little has been published on this
feature and its companion near 5.7 $\mu$m.

The weak 5.25 and 5.7 $\mu$m PAH features do not correspond to
fundamental vibrational frequencies ($\nu_{i}, \nu_{j}, \cdots$), but
are produced by overtones ($n\times\nu_{i}$), combinations ($\nu_{i} +
\nu_{j}$), and difference ($\nu_{i} - \nu_{j}$)bands of these
fundamental vibrations \citep{1989ApJ...345L..59A}. For example, the
strong CH out-of-plane (CH$_{\rm oop}$) fundamental bending vibration
($\nu_{\rm oop}$) for solo hydrogens produces the well known band at
11.2 $\mu$m (893 cm$^{\rm -1}$).  The overtone of this vibration,
$2\times\nu_{\rm oop}$, is expected to produce a much weaker feature
near 5.6 $\mu$m (1786 cm$^{\rm -1}$) and is seen to contribute to the
blue side of the broad 5.7 $\mu$m interstellar feature.  Likewise,
$\nu_{\rm oop}$ could combine with a CC stretching vibration
($\nu_{\rm oop} \pm \nu_{\rm CC}$), resulting in other weak features.

Here we present high quality ISO SWS \citep{1996A&A...315L..49D}
spectra from four astronomical sources that show these features. The
5.25 and 5.7 $\mu$m bands are analyzedin terms of overtone,
combination and difference frequencies using experimental and
theoretical PAH spectra. This manuscript is structured as follows. The
astronomical observations are presented in Sect. \ref{sec:data}, PAH
spectroscopy is described in Sect. \ref{sec:spectroscopy},
astrophysical implications are drawn in Sect. \ref{sec:astrophysical}
and a summary with conclusions is given in Sect. \ref{sec:summary}.

\section{The Astronomical emission features in the 5 - 6 $\mu$m (2000 to 1667 cm$^{\rm -1}$) region}
\label{sec:data}

\subsection{Observations}
\label{sec:observations}

A sample of high quality spectra from 15 sources, obtained by the
Short Wavelength Spectrograph (SWS) onboard ESA's Infrared Space
Observatory (ISO), were investigated for PAH band emission in the 5 -
6 $\mu$m region. Of this set, the spectra from four objects have
sufficient signal-to-noise in this wavelength range permit an in-depth
analysis.

The four sources considered here are: HD 44179 (`Red Rectangle'), NGC
7027, and two positions on the Orion ionization ridge, H2S1 and D5. HD
44179 is the central binary star system of a bipolar planetary nebula,
with one member being a post-AGB star. Both stars have a common
circumstellar disk \citep[e.g.][]{1996A&A...315L.245W}. NGC 7027 is a
compact ($\sim10^{16}$ cm) carbon-rich Planetary Nebula with a hot
White Dwarf at its center \citep{2000ApJ...539..783L}. The Orion Bar
spectra probe the Photon Dominated Region forming the interface
between the bright \ion{H}{2} region that is ionized by the Trapezium
stars, and the Orion Molecular cloud; H2S1 and D5 are two positions
within the bar. Table \ref{tab:astrometric} shows the journal of
observations and the available astrometric data.

\begin{deluxetable}{lcclllclc}
  \tablewidth{\textwidth}
  \tabletypesize{\scriptsize}
  \tablecaption{Journal of observations and available astrometric data. \label{tab:astrometric}}
  \tablehead{
  \colhead{Source} & \colhead{$\alpha$ (2000)} & \colhead{$\delta$} (2000)               & \colhead{TDT} & \colhead{Obs. mode} & \colhead{Sp.}  & \colhead{$G_{0}$}  & \colhead{Object Type} & \colhead{Ref.}    \\
  \colhead{}       & \colhead{[h m s]}         & \colhead{[$^{\circ}\ \arcmin\ \arcsec$]} & \colhead{}    & \colhead{(speed)}   & \colhead{Type} & \colhead{[Habing]} & \colhead{}            & \colhead{}
  }
  \startdata
  HD 44179         & 06 19 58.20               & -10 38 15.22                            & 70201801      & 01 (4)              & B8V            & $5\cdot10^{6}$     & Post-AGB star         & \tablenotemark{1} \\
  NGC 7027         & 21 07 01.70               & +42 14 09.10                            & 55800537      & 06                  & $2\ 10^{5}$ K  & $2\cdot10^{5}$     & PN                    & \tablenotemark{1} \\
  Orion Bar H2S1   & 05 35 20.31               & -05 25 19.99                            & 69501806      & 01 (4)              & O6             & $7\cdot10^{3}$     & \ion{H}{2} region     & \tablenotemark{2} \\
  Orion Bar D5     & 05 35 19.81               & -05 25 09.98                            & 83101507      & 01 (2)              & O6             & $5\cdot10^{4}$     & \ion{H}{2} region     & \tablenotemark{3} \\
  \enddata
  \tablerefs{(1) \cite{1996A&A...315L.369B}; (2) \cite{2001A&A...372..981V}; (3) \cite{2002A&A...390.1089P}}
\end{deluxetable}

\subsubsection{Data Reduction}
\label{sec:reduction}

\begin{figure}
  \plotone{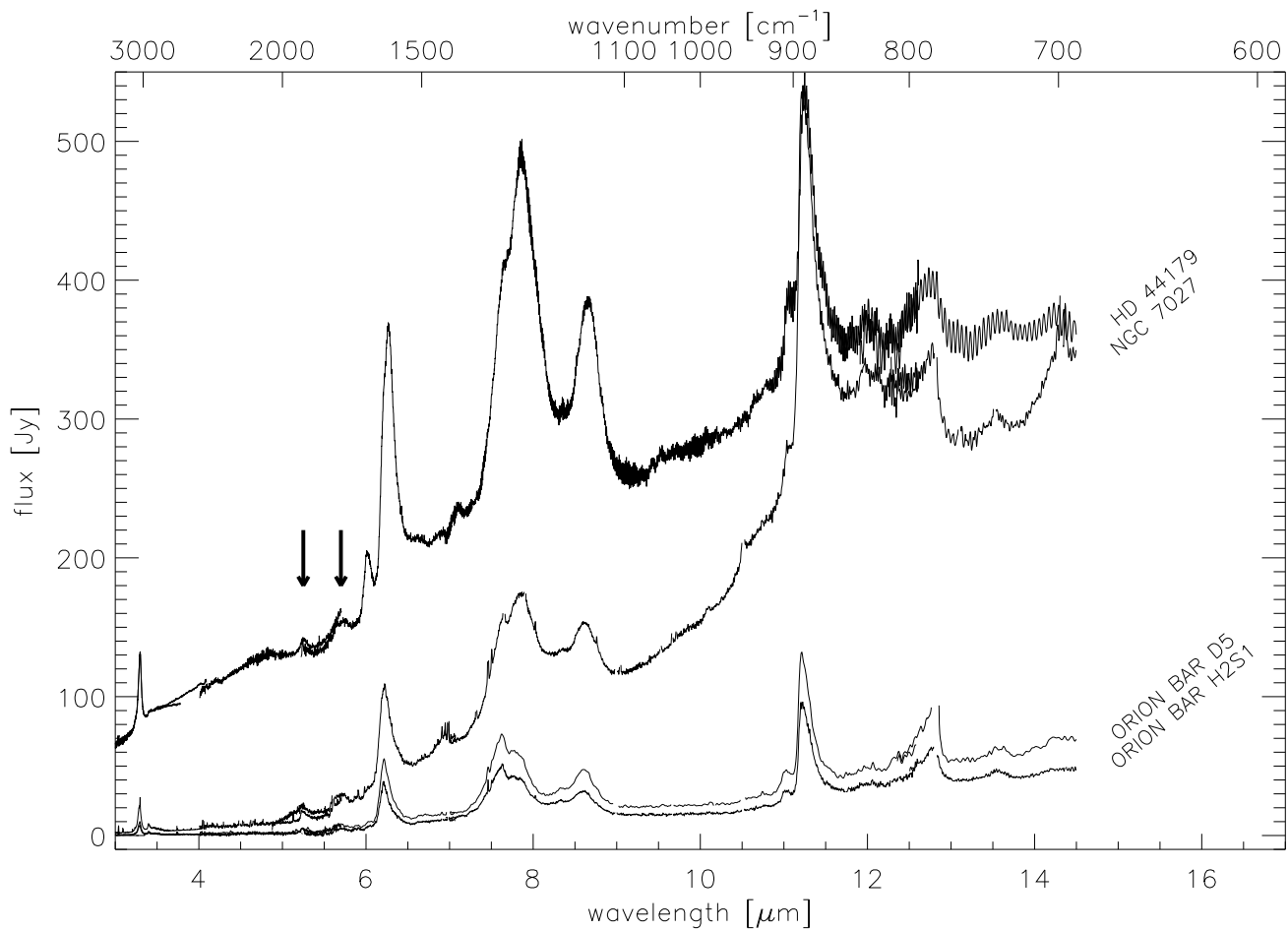}
  \caption{Extracted mid-IR spectra from 3 - 14.5 $\mu$m of the four sources considered here: HD 44179, NGC 7027 and the two positions along the Orion Bar. Spectra are shown at original resolution. For clarity the prominent atomic emission lines, e.g. at 5.61 ([\ion{Mg}{5}]), 10.51 ([\ion{S}{4}]) and 12.81 ([\ion{Ne}{2}]) $\mu$m, have been masked out. The arrows indicate the positions of the subtle 5.25 and 5.7 $\mu$m features. \label{fig:overview}}
\end{figure}

\begin{figure}
  \plotone{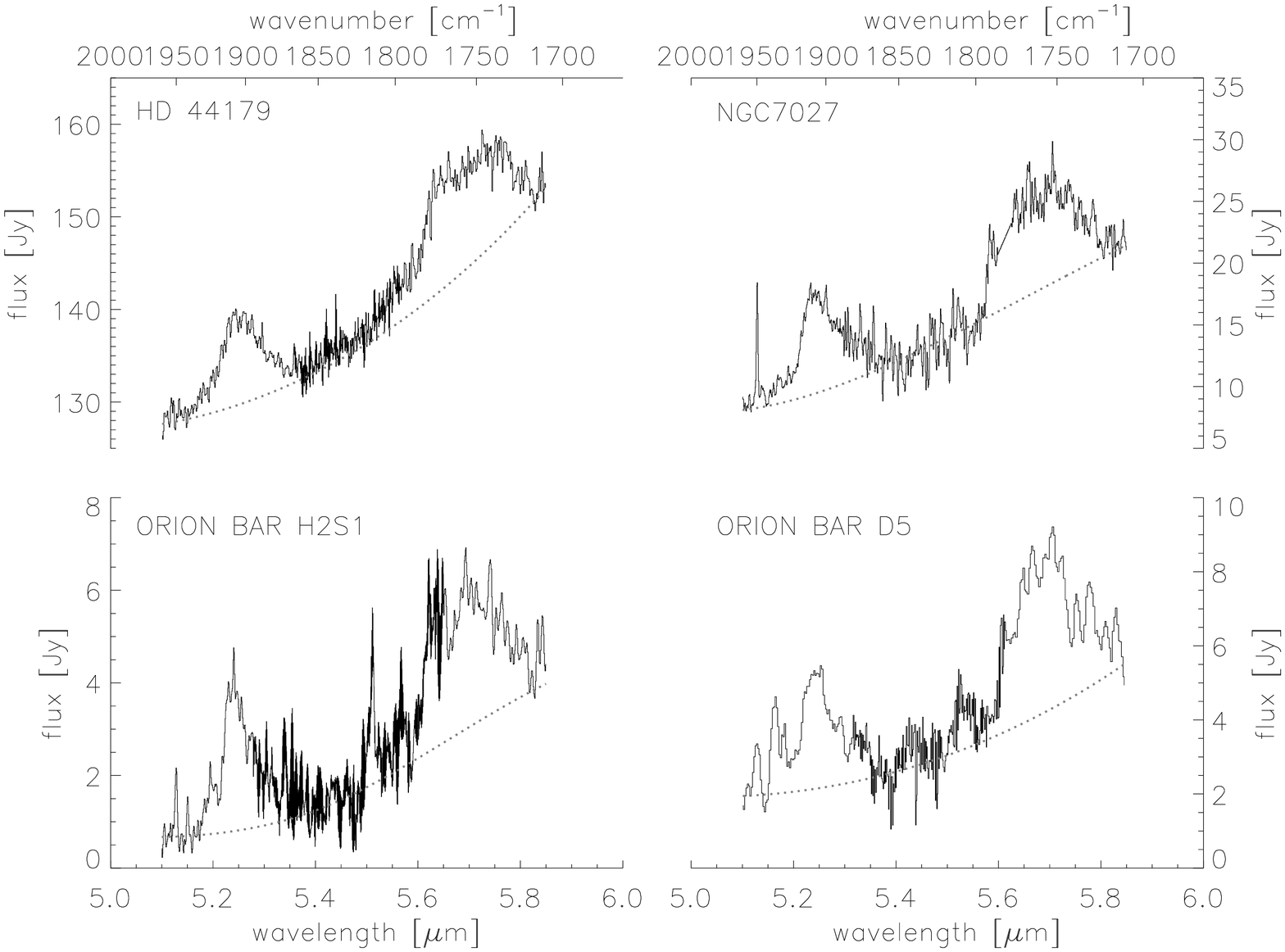}
  \caption{The fully reduced 5 - 6 $\mu$m spectra of the four objects considered. The are presented at original resolution. Also shown are the adopted continua. \label{fig:bands}}
\end{figure}

The data have been obtained using the AOT 01 full scan mode at various
speeds or the AOT 06 mode, providing spectra with resolving power of
400 to 1600 ($\lambda/\Delta\lambda$). The data were processed with
IA$^{\rm 3}$, the SWS Interactive Analysis package, using calibration
files and procedures equivalent with OLP version 6.0 or 10.1. A
detailed account of the reduction can be found in
\cite{2002A&A...390.1089P}. Fig. \ref{fig:overview} presents the raw
extracted 2.38 - 45.2 $\mu$m ISO SWS spectra of the four sources at
their original resolution. These data have not been defringed. The
OSIA\footnote[7]{OSIA is a joint development of the SWS
  consortium. Contributing institutes are SRON, MPE, KUL and the ESA
  Astrophysics Division} software package was drawn on to further
reduce and analyze the data between 5 and 6 $\mu$m. Further reduction
included bad data removal (e.g. glitches), edge truncation, splicing
and merging of detector bands. Uncertainties were estimated by
comparing the up and down scans, yielding a measure for the noise. For
HD 44179 a difference in absolute flux levels, not in shape, between
the up and down scan was found. The up scan was scaled such that the
mean of the up scan would match the mean of the down scan to
compensate this. This can be done because our interest lies only in
the uncertainty in relative flux level, since those trace the reality
of the shape of a feature. When required, the 5 - 6 $\mu$m spectra
were binned to a constant resolution of 800 to facilitate one-to-one
comparison. For Orion Bar position D5 this results in oversampling
since the original data were obtained at a resolution of about
400. Table \ref{tab:parameters} presents a subset of the data
reduction parameters and Fig. \ref{fig:bands} presents the fully
reduced 5 - 6 $\mu$m ISO SWS spectra of the four sources at their
original resolution.

\begin{deluxetable}{lcccl}
  \tablewidth{\textwidth}
  \tabletypesize{\footnotesize}
  \tablecaption{Data reduction parameters and connected wavelengths.\label{tab:parameters}}
  \tablehead{
  \colhead{Source}   & \colhead{OLP}             & Resolution      & \colhead{Continuum points} & \colhead{Integration bounds} \\
  \colhead{}         & \colhead{}                &                 & \colhead{[$\mu$m]}         & \colhead{$\mu$m}}      
  \startdata
    HD 44179         & 10.1                      & $\sim$800       & 5.017;5.494;5.836;5.922    & 5.160;5.355;5.580;5.830      \\
    NGC 7027         & 6.0/10.1\tablenotemark{1} & $\sim$1600/1000 & 5.025;5.079;5.832;5.938    & 5.150;5.375;5.570;5.805      \\
    Orion Bar H2S1   & 10.1                      & $\sim$800       & 5.102;5.415;5.825;5.969    & 5.170;5.374;5.590;5.820      \\
    Orion Bar D5     & 10.1                      & $\sim$400       & 5.057;5.382;5.472;5.943    & 5.150;5.380;5.560;5.845      \\
  \enddata
  \tablerefs{(1) The AOT 01 and AOT 06 data have been reduced with OLP version 6.0 and 10.1 respectively. The AOT 01 data is used in Fig. \ref{fig:overview} and the better quality AOT 06 data is presented in Fig. \ref{fig:bands} and used onwards.}
\end{deluxetable}

\subsection{Analysis}
\label{sec:analysis}

The spectra display a wealth of structure, including the familiar PAH
bands at 3.3, 6.2, `7.7', 8.6, 11.2 and 12.7 $\mu$m with their
underlying plateaus and sub-features. For clarity, several atomic
emission lines, have been masked out in
Fig. \ref{fig:overview}. Detailed analysis of the major features in
the spectra have been published elsewhere
\citep[e.g.][]{2001A&A...370.1030H, 2002A&A...390.1089P,
  2004ApJ...611..928V}. Here we focus on the PAH emission between 5 -
6 $\mu$m. As Fig. \ref{fig:overview} illustrates, the bands in this
region are weak compared to the major features.

To isolate the PAH band profiles a continuum is established for each
object. The continua used apply only to the 5 - 6 $\mu$m
region. Spline continua were constructed using connecting points
between 5 - 6 $\mu$m, straddling the 5.25 and 5.7 $\mu$m features. The
adopted continua are shown in Fig. \ref{fig:bands} and the connected
wavelengths are listed in Table \ref{tab:parameters}.

\begin{figure}
  \plotone{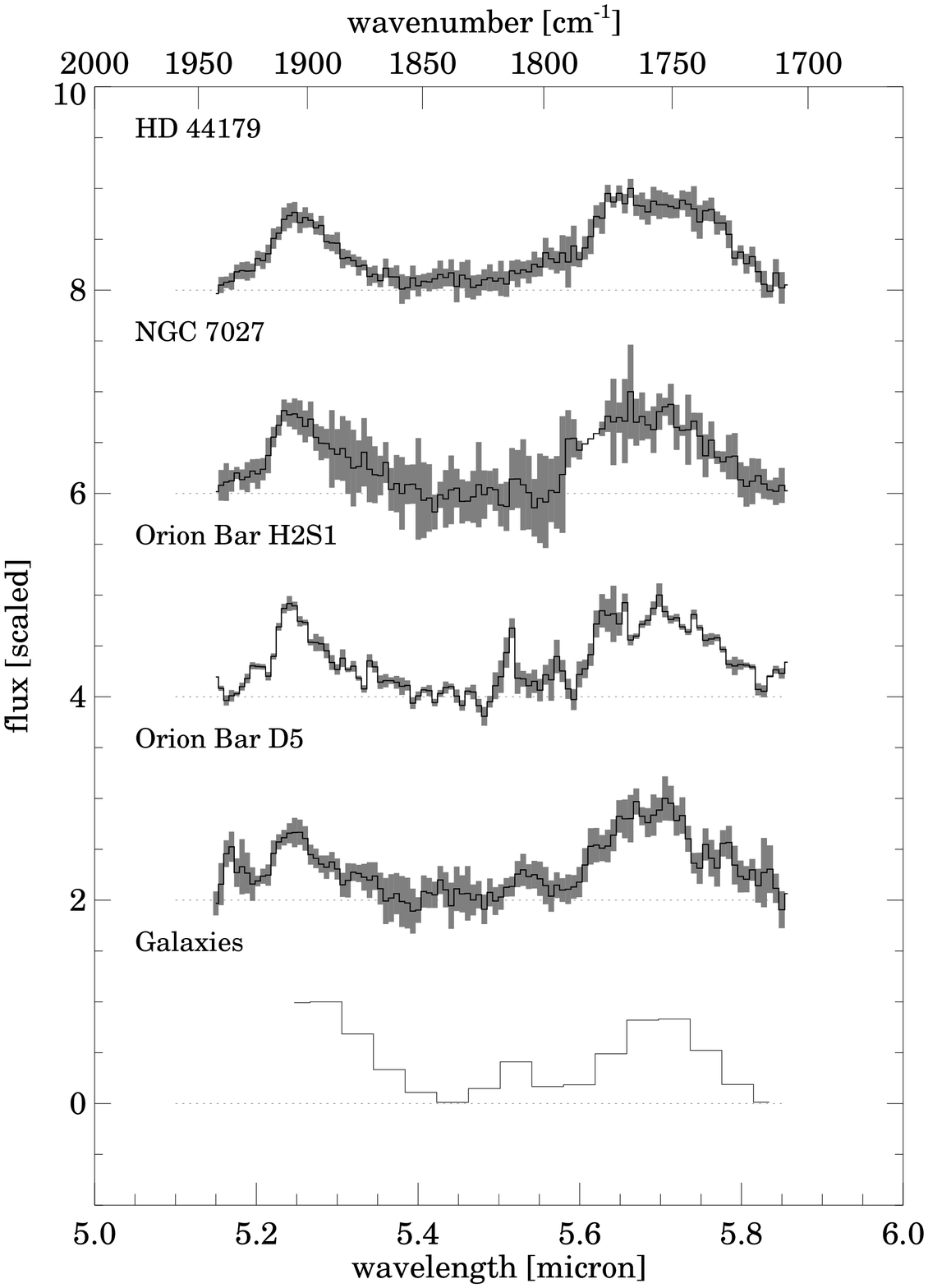}
  \caption{The 5.25 and 5.7 $\mu$m PAH features in HD 44179, NGC 7027 and the Orion Bar: positions H2S1 and D5. All spectra are presented at $\lambda/\Delta\lambda = 800$ to aid comparison. These spectra were produced using the continuum and emission line corrections described in Sect. \ref{sec:analysis}. To demonstrate the accessibility of the 5 - 6 $\mu$m region using Spitzer, the slightly redshifted average galaxy spectrum from \cite{2007ApJ...656..770S}, is also presented, see Sect. \ref{sec:astrophysical} for a discussion. \label{fig:compare}}
\end{figure}

Atomic emission lines can also contribute to the emission in this
region. For some of these objects narrow lines at 5.128 (1950 cm$^{\rm
  -1}$; \ion{H}{1} 6 - 10), 5.510 (1815 cm$^{\rm -1}$; H$_{\rm 2}$ 0
$\rightarrow$ 0 S(7)) and 5.61 $\mu$m (1783 cm$^{\rm -1}$;
[\ion{Mg}{5}]) are present. Special care was taken to remove the 5.61
$\mu$m [\ion{Mg}{5}] line in the spectrum of NGC 7027 where it is
blended with the 5.75 $\mu$m PAH feature. The resulting, continuum
subtracted spectra are presented at equal resolution in
Fig. \ref{fig:compare}. Two broad bands, centered roughly at 5.25 and
5.7 $\mu$m, are clearly present in all objects and are quite similar
within the uncertainty (systematic + noise). The 5.25 $\mu$m band is
asymmetric and sets off near 5.15 $\mu$m (1942 cm$^{\rm -1}$). For
both positions in the Orion Bar it also seems to start off with a
satellite feature, centered at about 5.2 $\mu$m (1923 cm$^{\rm -1}$)
and similar to those seen at 6.0 and 11.0 $\mu$m for the 6.2 and 11.2
$\mu$m bands, respectively. After a steep rise the profile peaks at
5.25 $\mu$m (1905 cm$^{\rm -1}$) and slowly declines to reach zero
intensity near 5.4 $\mu$m (1852 cm$^{\rm -1}$). This red wing
resembles that seen for the 3.4, 6.2 and 11.2 $\mu$m features
\citep{1996MNRAS.281L..25R}. Fitting the profiles at original
resolution with a single Gaussian results in a FWHM of about
$\sim$0.12 $\pm$ 0.01 $\mu$m ($\sim$42 $\pm$ 6.5 cm$^{\rm -1}$.

The 5.7 $\mu$m feature is broader, and starts its rise near 5.55
$\mu$m (1802 cm$^{\rm -1}$), just beyond the 5.51 $\mu$m (1815
cm$^{\rm -1}$) H$_{\rm 2}$ 0 $\rightarrow$ 0 S(7) line. The profile is
unusual from about 5.65 $\mu$m (1770 cm$^{\rm -1}$) to about 5.75
$\mu$m (1739 cm$^{\rm -1}$), where the feature seems flat-topped and
then drops to the continuum at 5.85 $\mu$m (1709 cm$^{\rm -1}$). Some
spectral structure is visible in on the flat-top, giving the
impression of a blended double-peaked feature. Fitting the profile
with a single Gaussian results in a FWHM of about 0.17 $\pm$ 0.02
$\mu$m ($\sim$51 $\pm$ 5.6 cm$^{\rm -1}$).

After emission line removal the bands are integrated in the
$\lambda-{\rm F_{\lambda}}$ domain, using a multiple Simpsons' rule,
yielding band strengths in W$\cdot$m$^{\rm -2}$. Uncertainties are
calculated from the data. Table \ref{tab:strengths} summarizes these
results and compares them with the band strengths of the dominant PAH
bands.

\begin{deluxetable}{llllllllll}
  \tablewidth{\textwidth}
  \tabletypesize{\tiny}
  \tablecaption{The integrated band strengths ($10^{-14}\ \mathrm{W}\ \mathrm{m}^{-2}$) of the PAH features between 3 - 13 $\mu$m of HD 44179, NGC 7027 and the Orion Bar: positions H2S1 and D5. Uncertainties are given in parentheses in percentages.\label{tab:strengths}}
  \tablehead{
  \colhead{Source}   & \colhead{3.3\tablenotemark{1}}  & \colhead{5.25}     & \colhead{5.75}     & \colhead{6.2\tablenotemark{1}} & \colhead{7.6\tablenotemark{1,a}} & \colhead{7.8\tablenotemark{1,a}} & \colhead{8.6\tablenotemark{1}}  & \colhead{11.2\tablenotemark{2}}     & \colhead{12.7}              \\
  \colhead{}         & \colhead{[$\mu$m]}              & \colhead{[$\mu$m]} & \colhead{[$\mu$m]} & \colhead{[$\mu$m]}             & \colhead{[$\mu$m]}               & \colhead{[$\mu$m]}               & \colhead{[$\mu$m]}              & \colhead{[$\mu$m]}                  & \colhead{[$\mu$m]}}
  \startdata
  HD 44179           & 69 (1.4\%)                      & 11 (1.5\%)         & 19  (1.3\%)         & 280 (10\%)                     & 121 (10\%)                     & 290 (10\%)                        & 97 (10\%)                       & 113 (10\%)                          & 26 (5.4\%)\tablenotemark{3}\\
  NGC 7027           & 24 (1.7\%)                      & 9.1 (3.3\%)        & 13  (7.0\%)         & 95 (10\%)                      & 40  (10\%)                     & 63  (10\%)                        & 28 (10\%)                       & 130 (10\%)                          & 36 (6.9\%)\tablenotemark{3}\\ 
  Orion Bar H2S1     & 8.2 (1.6\%)                     & 2.9 (1.6\%)        & 4.4 (1.7\%)         & 34 (10\%)                      & 32  (10\%)                     & 22  (10\%)                        & 13 (10\%)                       & 29  (10\%)                          & 13 (10\%)\tablenotemark{2} \\ 
  Orion Bar D5       & 11 (0.55\%)                     & 3.7 (3.2\%)        & 6.1 (2.3\%)         & 12 (10\%)                      & 40  (10\%)                     & 36  (10\%)                        & 24 (10\%)                       & 40  (10\%)                          & 17 (7.6\%)                 \\
  \enddata
  \tablenotetext{a}{7.6 and 7.8 $\mu$m bands have been separated using Gaussians}
  \tablerefs{(1) \cite{2002A&A...390.1089P}; (2) \cite{2004ApJ...611..928V}; (3) \cite{2001A&A...370.1030H}}
\end{deluxetable}

\section{PAH Spectroscopy in the 5 to 6 $\mu$m (2000 to 1667 cm$^{\rm -1}$) Region}
\label{sec:spectroscopy}

\subsection{Experimental Studies}
\label{sec:experimental}

The matrix isolation infrared spectroscopy techniques employed in this
study have been described in detail previously
\citep{1995JPC...99..3033H, 1995JPC...99..8978H, 1998JPC..102...329H}
and will be summarized here only briefly.

Matrix isolated PAH samples were prepared by vapor co-deposition of
the species of interest with an overabundance of argon onto a 14K CsI
window suspended in a high-vacuum chamber (p $<10^{-8}$ Torr). The
samples were vaporized from heated Pyrex tubes while argon was
admitted through an adjacent length of copper tubing, cooled by liquid
nitrogen. The deposition temperatures, which are dependent on size and
structure as well as the sources of the individual PAHs, are given in
the references summarized in Table 1 of \cite{2005ASR...36...156M}.

Spectra from 1.7 - 20 $\mu$m (6000 to 500 cm$^{\rm -1}$) were measured
on either a Nicolet 740 or a Digi-Lab Excalibur FTS 4000 FTIR
spectrometer using a KBr beam splitter and a liquid nitrogen cooled
MCT detector. Each spectrum represents a co-addition of between 500
and 1024 scans at a resolution of 0.5 cm$^{\rm -1}$. The number of
scans was chosen to optimize both the signal-to-noise as well as time
requirements for each experiment.

\begin{figure}
  \plotone{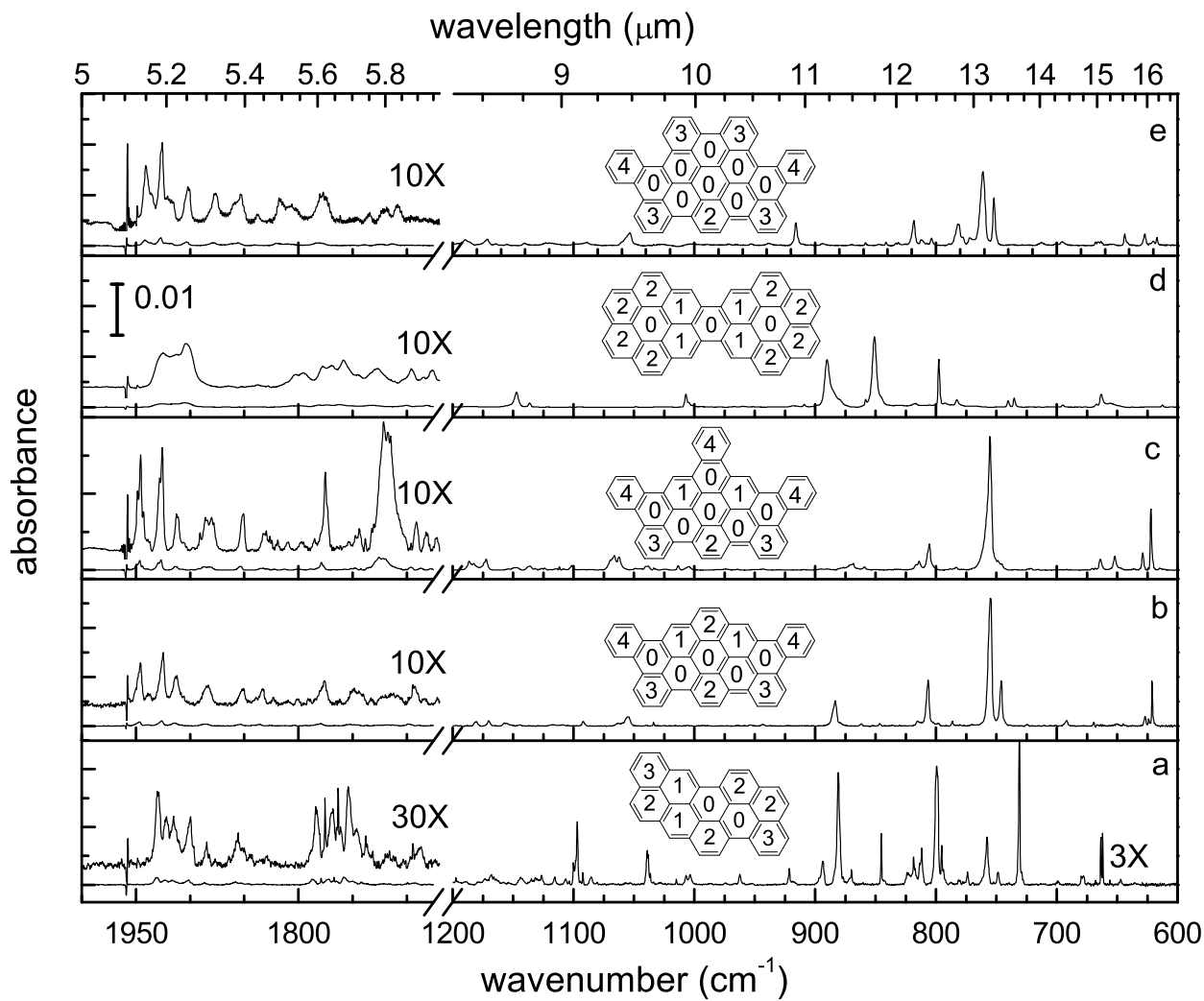}
  \caption{Matrix isolated spectra for the CH$_{\rm oop}$ region (10 - 15 $\mu$m) as well as the overtone and combination band region (5 - 6 $\mu$m) for various neutral PAHs containing solo, duo, trio and quartet hydrogens. The number of adjacent hydrogen atoms per ring are indicated in the structure.\label{fig:combination}}
\end{figure}

Figure \ref{fig:combination} shows the 5 to 16.5 $\mu$m spectra of
several matrix isolated neutral PAHs. These spectra are typical of all
spectra in the \emph{Ames} collection of neutral PAH IR spectra and
serve to illustrate that all PAHs have weak bands in the 5 to 6 $\mu$m
region. For presentation purposes only, some of the data has been
baseline corrected using the Digi-Lab data analysis software
package. No further data reduction was necessary. All numerical values
were obtained from the original (unaltered) data.

\subsection{Theoretical}
\label{sec:theoretical}

Previous work \citep[e.g.][]{1997Spect..53.1225B, 2002ApJ...564..782B,
  2003...107..1486M, 2005ApJ...632..316H, 2008ApJ...678..316B} have
shown that density functional theory (DFT) calculations performed
using the hybrid \citep{becke:5648} B3LYP \citep{1994JPC...98..11623S}
approach in conjunction with the 4-31G basis set \citep{frisch:3265}
yield scaled harmonic frequencies that are in excellent agreement with
results obtained from matrix isolation experiments. In this work, we
expand our previous work by explicitiy considering anharmonic effects
(fundamentals, overtones, combination, and difference bands), which
are computed by doing numerical differentiation along the normal modes
\citep{barone:014108} after the geometries are fully optimized and the
harmonic frequencies are computed using analytic second
derivatives. All of the DFT calculations are performed using the
Gaussian 03 program system \citep{g03}. The vibrational modes are
illustrated using Molekel\footnote[8]{MOLEKEL 4.3, P. Fl\"ukiger,
  H.P. L\"uthi, S. Portmann, J. Weber, Swiss Center for Scientific
  Computing, Manno (Switzerland), 2000-2002}
\citep{2000MOL..........P}.

As discussed previously, the DFT results differ systematically from
experiment and hence need to be scaled
\citep{1997Spect..53.1225B}. The scaling factor depends on the level
of theory (the correlation treatment and basis set), and it is common
that CH stretches will have a different scaling factor from the other
modes because of the large anharmonic effects for hydrogen stretching
motions. In this work we use the small 4-31G set that has been used
successfully in previously and the significantly larger and more
precise, but also more time consuming, 6-311G** basis set
\citep{frisch:3265}. The larger 6-311G** basis set was used for
naphthalene (C$_{\rm 10}$H$_{\rm 8}$) because of its small size. The
results from the large basis set allow us to calibrate the smaller
basis set that is required for the larger molecules. We use the matrix
isolation experiments of \cite{1994JPC...98..4243H} for the
calibration of neutral naphthalene. While gas-phase data exists, we
use the matrix isolation experiments to allow a direct comparison
between our experiments and our computed values. In addition, it has
been shown by e.g. \cite{2000ApJ...542..404O} and
\cite{2003ssac.proc..251A} that ther is a very good agreement between
the gas phase and matrix isolation results.

Comparison of the computed fundamental frequencies of neutral
naphthalene, using the 6-311G** basis set, shows that a scale factor
of 0.998 brings the computed non-CH stretching modes (i.e. all the
bands excluding the CH stretching modes) into good agreement with
experiment. This comparison shows an average absolute error of 2.6
cm$^{\rm -1}$ and a maximum absolute error of 9 cm$^{\rm -1}$ between
experiment and theory. A scale factor of 1.005 brings the computed CH
stretching modes into good agreement (a maximum error of 7 cm$^{\rm
  -1}$) with experiment. This accordance with experiment is consistent
with recent work by \cite{CaneE._jp071610p}, who find good agreement
between theory and experiment using a hybrid functional and a large
basis set (B971/TZ2P). For the smaller 4-31G basis set we find a scale
factor of 0.976 for the non-CH stretching fundamentals and 1.007 for
the CH stretching fundamentals. The maximum absolute error in the
non-CH stretching modes is 19 cm$^{\rm -1}$, a value greater than that
when using the larger basis set, but still sufficiently small to allow
a qualitative understanding using the small basis set. It is
interesting to note that two scale factors are needed for the 4-31G
fundamentals while only one scale factor is needed for the 4-31G
harmonics.

Since the scale factors for the 6-311G** basis set are very close to
1, we do not use any scaling for the overtone, combination, or
difference bands. For the 4-31G basis set, the overtone and
combination bands are scaled by 0.976, the scale factor for the non-CH
stretching modes, because the CH stretches are not involved in the
overtone or combination bands of interest. For the difference bands,
the non-CH stretching fundamentals are scaled by 0.976 while the CH
stretching fundamentals are scaled by 1.007. While the level of theory
used is expected to yield accurate band positions for the overtones,
combination, and difference bands, the current version of Gaussian
does not determine intensities for these bands. However, as discussed
in Sec. \ref{sec:theory}, selection rules allow the exclusion of some
combination and difference bands as possible contributors to the
observed features. In addition to excluding some bands, we speculate
on which allowed bands will be the strongest.  In the harmonic
approximation, the second derivative of the dipole moment is
responsible for the intensity of the first overtone, but in the real
world, anharmonicity means that the overtones will have some
contributions from the first derivative. This implies that the
strongest overtone, combination, and difference bands will commonly
involve the strongest fundamentals.

\begin{figure}
  \plotone{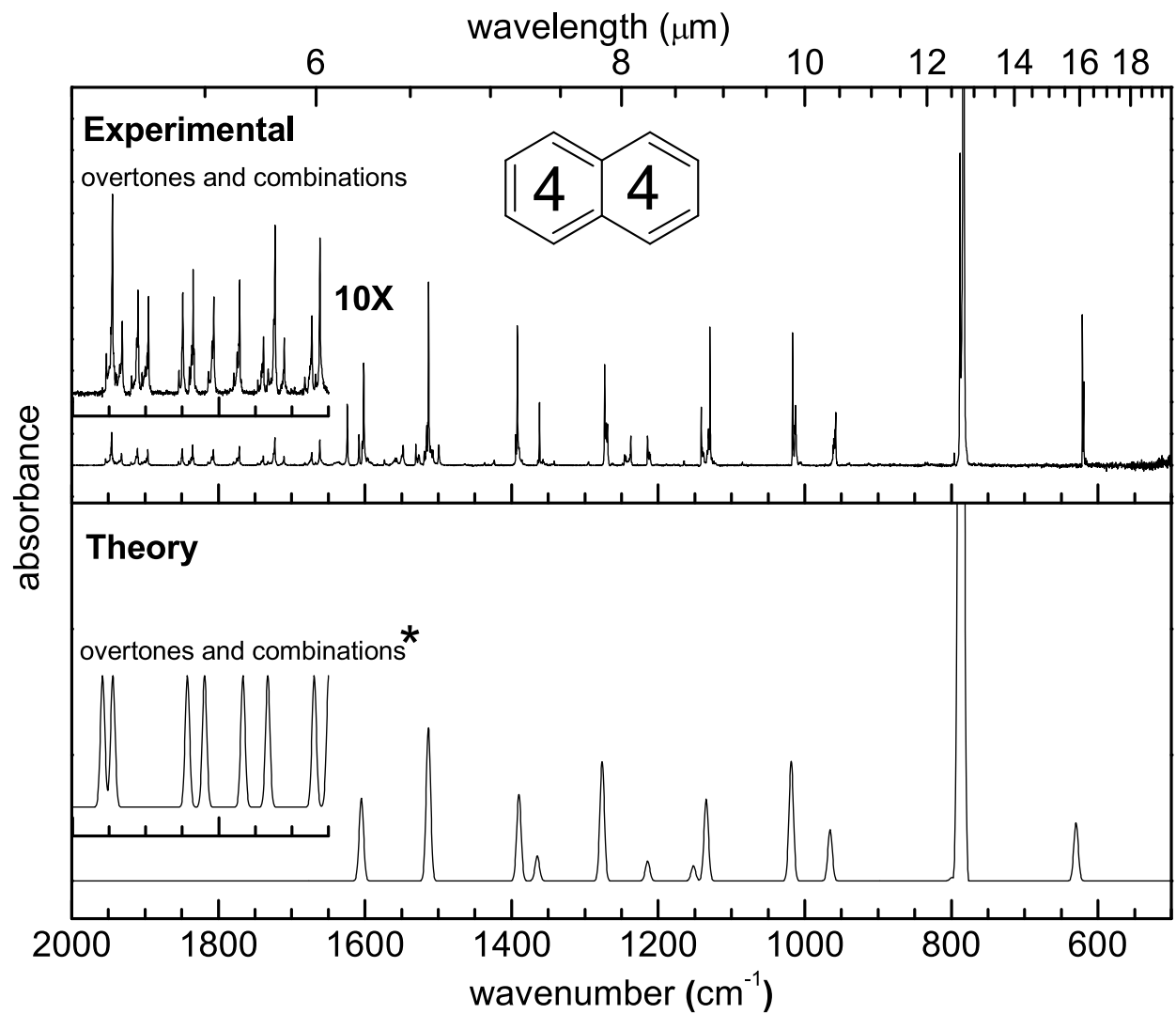}
  \caption{Comparison between the B3LYP-level calculated and experimentally obtained spectra of naphthalene from 5 to 19 $\mu$m, showing their outstanding agreement for fundamentals. The agreement for the overtone and combination bands is reasonable. *Note that since the theoretical results for the overtone and combination bands do not calculate intensities. Therefore, the bands were derived from fundamentals with an intensity $>$10 km/mole and given the same intensity to allow for comparison with the experimental frequencies. \label{fig:match}}
\end{figure}

Figure \ref{fig:match} compares the 5 to 19 $\mu$m (2000 to 530
cm$^{\rm -1}$) spectrum of neutral naphthalene computed using density
functional theory at the B3LYP level to the spectrum of matrix
isolated neutral naphthalene (C$_{\rm 10}$H$_{\rm 8}$) from
\cite{1994JPC...98..4243H}. This figure demonstrates the excellent
agreement between the theoretically computed and experimentally
measured mid-IR spectrum for fundamentals and the reasonable agreement
in the positions of the calculated overtone and combination
bands. Note that difference bands only arise from emission processes
and are thus not found in the experimental spectra.

\subsection{The Origin of the PAH Bands in the 5 to 6 $\mu$m (2000 to 1667 cm$^{\rm -1}$) range}
\label{sec:origin}

The experimental spectra presented in Fig. \ref{fig:combination}
illustrate what has long been known, PAHs show weak spectral structure
in the 5 to 6 $\mu$m (2000 to 1667 cm$^{\rm -1}$) region. This
structure has been attributed to overtones and combination bands based
on spectroscopic studies of small aromatic molecules
\citep{1951Ana.....23...709}. This work is summarized below, followed
by a detailed analysis of the PAH vibrations that produce bands in the
5 to 6 $\mu$m region for large PAHs, species comparable in size to the
astronomical PAHs.

\cite{1951Ana.....23...709} present some examples of the distinctive
band patterns in the 5 to 6 $\mu$m region produced by different
substituted benzenes. The different substitution patterns force
different H adjacencies and this has been found to determine the
characteristic band patterns in this region. This behavior strongly
suggests that fundamental vibrations associated with hydrogen are the
primary carriers of this spectroscopic structure, but these different
substitution patterns could also induce IR activity in transitions
involving CC vibrations. The central role of hydrogen in producing
these features is confirmed when one compares the 5 to 6 $\mu$m region
of the spectra of normal PAHs (C$_{\rm x}$H$_{\rm y}$) with the same
region in the spectra of fully deuterated PAHs (PADs; C$_{\rm
  x}$D$_{\rm y}$). The absence of any activity in the 5 to 6 $\mu$m
region of the deuterated PAH spectra shows that hydrogen is essential
in producing these features. The band patterns in the 5 to 6 $\mu$m
region reflect hydrogen adjacency because, as discussed below, they
involve strong CH vibrations whose fundamental frequencies are
determined by their adjacency class.

\subsubsection{Experimental}
\label{sec:exp}

The 5 to 6 $\mu$m region of the astronomical emission spectrum was
initially thought to be dominated by simple combinations and overtones
of the fundamental CH$_{\rm oop}$ modes \citep{1960tisocm.book.....N}.
Thus our analysis begins by separating out the contributions of the
CH$_{\rm oop}$ modes. The CH$_{\rm oop}$ region, which falls between
10.5 and 15 $\mu$m, includes contributions of the solo (non-adjacent
hydrogens), duo (doubly adjacent hydrogens), trio (triply adjacent
hydrogens) and quartet (quadruply adjacent hydrogens) hydrogens
present on a PAH \citep[e.g.][]{1999ApJ...516L..41H}. Each of these
type of vibrations falls within specific wavelength regions. Given the
segregation of the CH$_{\rm oop}$ vibrations in PAH spectra one would
expect combination modes arising from these fundamental vibrations to
be similarly grouped (i.e. quartets furthest to the red and solos
furthest to the blue), if the various CH$_{\rm oop}$ modes combined
with the same vibrational modes. To better understand this region, we
have analyzed the spectra of over 35 neutral PAHs of various sizes and
structures and have found that this is not necessarily the case. While
several features between 5 and 6 $\mu$m appear to be the result of
solo, duo, trio and quartet hydrogens, their order does not follow
that observed in the CH$_{\rm oop}$ region.

The CH$_{\rm oop}$ spectra of several matrix isolated PAHs are shown
in conjunction with their respective overtone and combination regions
in Fig. \ref{fig:combination}. As one can easily see, the CH$_{\rm
  oop}$ region varies considerably with respect to the various PAH
structures, reflecting the variability in the solo, duo trio and
quartet hydrogens from one PAH to the next. Likewise there is
considerable variability in the overtone and combination region (5.1
to 5.8 $\mu$m region). Close inspection of Fig. \ref{fig:combination}
reveals that molecules {\bf b}, {\bf c}, and {\bf e} contain a large
set of bands spanning the full wavelength region, while molecules a
and d have a pronounced break in the overtone and combination bands
between the 5.25 and 5.7 $\mu$m regions. The major structural
difference in these PAHs is the presence of quartet hydrogens in
molecules {\bf b}, {\bf c}, and {\bf e}. The presence of quartet
hydrogens in a PAH molecule produces a complex of bands spanning the
entire 5 and 6 $\mu$m region, making them unsuitable as responsible
for the astronomical 5.25 and 5.7 $\mu$m features.

\begin{figure}
  \plotone{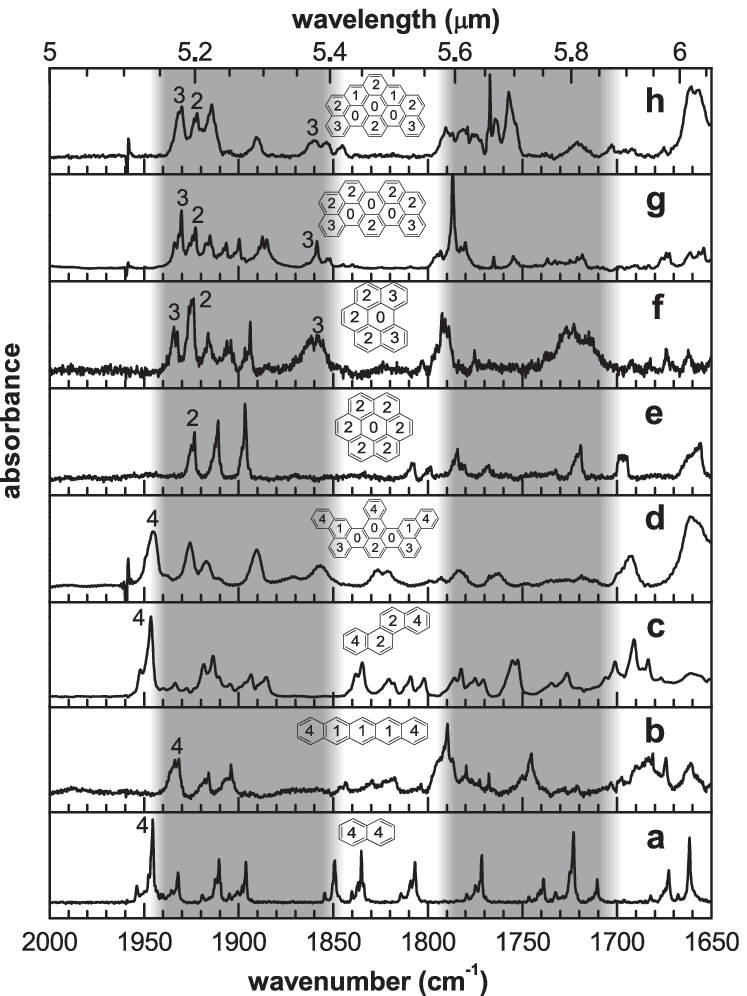}
  \caption{Matrix isolated spectra of (a) naphthalene, (b) pentacene, (c) chrysene, (d) large, uncondensed PAH containing solo, due, trio and quartet hydrogens, (e) coronene, (f) benzo[g,h,i]perylene, (g) large PAH containing only duo and trio hydrogens and (h) large PAH containing duo, trio and solo hydrogens. Bands identified as involving quartet hydrogens are labelled with a 4, trio hydrogens a 3, and duos a 2. The shaded areas show the band positions of the 5.25 and 5.7 $\mu$m astronomical features. \label{fig:adjacency}}
\end{figure}

Ignoring, for the moment, the fundamental vibrational modes as carrier
of the overtone and combination bands, several additional interesting
trends appear in the 5 to 6 $\mu$m region which can be linked to the
PAH structure. The PAH spectra presented in Fig. \ref{fig:adjacency}
will assist in identifying these trends. The positions of the
astronomical 5.25 and 5.7 $\mu$m have been indicated by the shaded
areas, with on both sides an uncertainty of $\sim$10 cm$^{\rm -1}$ to
account for cold absorption versus hot emission
\citep{1995A&A...299..835J}. The first trend to note is that all the
matrix-isolated PAHs in Fig. \ref{fig:adjacency}, along with all the
PAHs investigated by this lab to-date, exhibit bands in the 5.25 and
5.7 $\mu$m regions. The quartet containing PAHs, in addition to
exhibiting bands in undesirable positions, also tend to have bands to
the blue and red of the positions occupied by the 5.25 and 5.7 $\mu$m
astronomical features. For instance, in Fig. \ref{fig:adjacency}a the
bands extend further to the blue of the 5.25 $\mu$m area than the
bands visible in the PAHs shown in panels {\bf e}, {\bf f}, {\bf g}
and {\bf h} of Fig. \ref{fig:adjacency}, which contain only solo, duo
and trio hydrogens. Furthermore, quartets that are part of a `bay'
region in PAHs, such as found in panels {\bf c} or {\bf d} of
Fig. \ref{fig:adjacency}, tend to produce bands at wavelengths shorter
than those of less hindered quartets, such as shown in panels {\bf a}
and {\bf b} of Fig. \ref{fig:adjacency}. Typically the bay region
quartets result in a band around 5.128 $\mu$m (1950 cm$^{\rm -1}$)
versus the 5.14 $\mu$m (1940 cm$^{\rm -1}$) position of the less
hindered quartet. Thus PAHs containing quartet hydrogens will exhibit
bands outside the area occupied by the 5.25 and 5.7 $\mu$m
astronomical features.

Additional analysis reveals that PAHs with duo hydrogens tend to
exhibit a band around 5.19 $\mu$m (1925 cm$^{\rm -1}$), see panel {\bf
  e} in Fig. \ref{fig:adjacency}. The presence of trio hydrogens, in
addition to duos, tends to produce another feature to the blue of the
duo band around $\sim$5.18 $\mu$m (1930 cm$^{\rm -1}$). This is
clearly illustrated in the spectra shown in panels {\bf e}, {\bf f},
{\bf g} and {\bf h} of Fig. \ref{fig:adjacency}, where the trios,
labeled as 3, in panel {\bf f}, {\bf g} and {\bf h} of
Fig. \ref{fig:adjacency} produce a band further to the blue of the duo
band, labeled as 2, in panel {\bf e} of
Fig. \ref{fig:adjacency}. However, in the absence of duo hydrogens,
the trios produce a single feature near the position of the lone duo
hydrogens. Finally the presence of hydrogen trios also tends to
produce a feature just to the red of the 5.25 $\mu$m band area
($\sim$5.39 $\mu$m or 1855 cm$^{\rm -1}$). This feature is readily
observed (labeled with a 3) in the matrix isolated spectrum shown in
panel {\bf f}, as well as in the spectra shown in panels {\bf g} and
{\bf h} of Fig. \ref{fig:adjacency}. Although all of the PAHs with
trio hydrogens exhibited this feature, the intensity of this band
depends on the specific PAH structure. Thus, PAHs containing only duo
and trio hydrogens match the 5.25 $\mu$m astronomical feature well,
but PAHs containing trio hydrogens match only marginally due the
presence of a band around 5.39 $\mu$m. This implies a limited
concentration of PAHs containing trio hydrogens in astronomical
objects.

The matrix isolated PAH absorption spectra in the 5.7 $\mu$m
astronomical band region, running from $\sim$5.5 to 5.8 $\mu$m,
exhibit considerably more variability than the 5.25 $\mu$m
region. Just as in the 5.25 $\mu$m region, every PAH investigated
shows bands in the 5.7 $\mu$m region.  However there seems to be a
greater degree of segregation concerning the absorptions in this
region. While the presence of quartets hydrogens produce bands in the
5.7 $\mu$m region, their presence also tends to produce bands to the
blue and red of the region. This is exemplified in the spectra shown
in panels {\bf a}, {\bf b} and {\bf c} of
Fig. \ref{fig:adjacency}. Although PAHs containing only duo and trio
hydrogens match the 5.7 $\mu$m feature very well, their bands tend to
cluster around 5.6 $\mu$m (1780 cm$^{\rm -1}$), whereas the presence
of solos extends this structure towards 5.8 $\mu$m (1724 cm$^{\rm
  -1}$). This is beautifully illustrated in the PAH spectra shown in
panels {\bf f}, {\bf g} and {\bf h} of Fig. \ref{fig:adjacency}. The
PAHs producing the spectra in panels {\bf f} and {\bf g} of
Fig. \ref{fig:adjacency} contain only doubly and triply adjacent
hydrogens, while the top spectrum is of a similar PAH but containing
solo hydrogens. As is easily seen in Fig. \ref{fig:adjacency},
although PAHs containing duo and trio hydrogens can contribute to the
5.7 $\mu$m astronomical feature, solo hydrogens are required to match
the full 5.7 $\mu$m feature.

In addition to possessing bands to the blue of the 5.25 $\mu$m feature
and bands to the blue and red of the 5.7 $\mu$m feature, PAHs
containing quartet hydrogens exhibit an additional trait which
prohibit them from effectively matching the 5.25 and 5.7 $\mu$m
astronomical regions. This concerns the region between the 5.25 and
5.7 $\mu$m astronomical features. As shown in panel {\bf a}, {\bf c}
and {\bf d} of Fig. \ref{fig:adjacency}, PAHs containing quartets
exhibit more intense activity in the area \emph{between} the 5.25 and
5.7 $\mu$m features than PAHs containing only solos, duos and
trios. In fact, as all the spectra in Figs. \ref{fig:combination} and
\ref{fig:adjacency} reveal, PAHs containing quartets seem to exhibit a
continuum of bands throughout the region that includes the 5.25 and
5.7 $\mu$m features. This is exemplified in the PAH spectrum shown in
panel {\bf d} of Fig. \ref{fig:adjacency}. Although it is a fairly
large PAH, its open structure contains solo, duo, trio and quartet
hydrogens. Likewise its spectrum exhibits a continuum of bands
throughout the region containing the 5.25 and 5.7 $\mu$m features.

Two final notes should be made regarding PAH structure and the
overtone and combination region. First, \emph{not all of the
  experimentally measured features in this region can be traced back
  to the strongest solo, duo, trio and quartet hydrogen bands}. This
precludes a one-to-one assignment with the theoretical results for all
the bands presented in the next section. This region also contains
bands around 1903 cm$^{\rm -1}$ (5.25 $\mu$m; $\pm$5 cm$^{\rm -1}$)
and 1915 cm$^{\rm -1}$ (5.22 $\mu$m; $\pm$5 cm$^{\rm -1}$) which
appear in the PAH spectra collected, regardless of structure.  In a
few instances either the 1903 or 1915 cm$^{\rm -1}$ band will appear
by itself, but in most instances both bands are present. Due to the
appearance of at least one of these bands in all of the PAH spectra,
we were unable to associate them with a particular structural
feature. However, the absence of any bands between 5 - 6 $\mu$m in
fully deuterated PAH spectra indicate that hydrogen motion is
involved, obviously through CH in-plane (CH$_{\rm ip}$) modes. Second,
while difference bands only arise from emission processes and are thus
not found in the absorption spectra presented here, they will likely
contribute to the astronomically observed features.

In summary, the experimental evidence suggests that the astronomical
5.25 and 5.7 $\mu$m features can be accommodated by the overtone and
combination bands resulting from PAHs containing trio, duo and solo
hydrogens. The presence of a band around 5.39 $\mu$m in PAHs
containing trio hydrogens implies a limit of the concentration of this
particular PAH structure in the astronomical PAH population. The
presence of quartet hydrogens produces undesirable features in the
spectrum. The 5.7 $\mu$m region exhibits a greater variability in
structure than the 5.25 $\mu$m. This greater variability is the result
of PAHs with only duo and trio hydrogens producing absorption features
which cluster on the blue side of the 5.7 $\mu$m region, whereas the
presence of solo hydrogens is necessary to produce features towards
the red side of this region. Such variability could be utilized as a
tracer of PAH structures in astronomical environments.

\subsubsection{Theoretical Studies}
\label{sec:theory}

Computational spectra permit an extensive analysis of the nature of
the transitions that produce features in the 5 - 6 $\mu$m region,
including difference bands which are inaccessible through experimental
absorption spectroscopy. Figure \ref{fig:match} compares the spectrum
of neutral naphthalene computed using density functional theory at the
B3LYP level to the spectrum of matrix isolated neutral naphthalene
(C$_{\rm 10}$H$_{\rm 8}$). This figure illustrates two important
points. First it demonstrates the excellent agreement between the
experimentally measured and theoretically computed \emph{fundamental}
transitions that produce the mid-IR spectrum of naphthalene. Second,
this level of theory does not predict any fundamental PAH vibrations
between 5 and 6 $\mu$m, adding further support of their assignment to
combination and overtone types of bands.

\begin{figure}
  \plotone{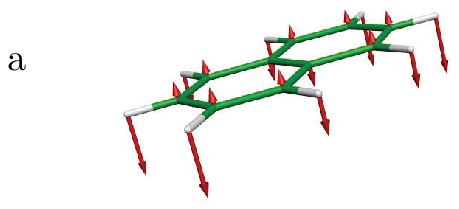}
  \caption{Example of one fundamental CH vibrational mode illustrative of the combination and difference bands of neutral naphthalene listed in Table \ref{tab:naphthalene}. The carbon skeleton is depicted in green, the hydrogen atoms in white and the vibrational motion of the molecule is indicated by the red vector arrows. Visualizations for all modes are available in the electronic edition of the Journal. \label{fig:neutral}}
\end{figure}

We now consider neutral naphthalene using the {6-311G**} basis
set. The results are summarized in the top two sections of Table
\ref{tab:naphthalene}. There are no overtones between 5 - 6 $\mu$m
that are derived from fundamentals with intensity greater than 10
km/mole. Therefore we conclude that overtones are not responsible for
the bands between 5 and 6 $\mu$m in the spectrum of
naphthalene. However, naphthalene is a special case since it has only
quartet hydrogens and, as shown in Table \ref{tab:coronene} for
coronene, overtones can produce important bands in this region. Now,
consider combination and difference bands, not overtones. Because of
the high symmetry in naphthalene, dipole-allowed combination and
difference bands can only arise from a forbidden mode interacting with
an allowed mode. In addition to this rigorous selection rule, we
presume that the stronger bands will result from combination and
differences in which the allowed band is intense
(Sect. \ref{sec:theoretical}). The combination and difference bands of
the most interest in the 5 - 6 $\mu$m range are also summarized in
Table \ref{tab:naphthalene}. All of the computed combination bands in
neutral naphthalene involve the strong CH$_{\rm oop}$ bending mode
(B$_{\rm 3u}$) at 12.703 $\mu$m. The vibration producing this strong
band couples with forbidden in-plane and out-of-plane CH bending
vibrations to produce the combination bands. The fundamental modes are
visualized in Fig. \ref{fig:neutral}. Hence, since the only strong
bands in neutral naphthalene are the CH$_{\rm oop}$ bend and the CH
stretch, and the higher frequency CH stretch cannot contribute to the
combination bands in the 5 - 6 $\mu$m region, the only bands in the
correct region that can couple with the strong B$_{\rm 3u}$ CH$_{\rm
  oop}$ band to produce features in the correct region are other CH
bends. Given that difference bands will involve a higher frequency
mode and that the only strong, allowed higher frequency fundamental is
the allowed CH stretch, all difference modes falling between 5 and 6
$\mu$m must involve coupling of the allowed CH stretch with
\emph{forbidden} CH$_{\rm ip}$ bending vibrations.

\begin{deluxetable}{c@{\hspace{5mm}}ccccllll}
  \tablewidth{\textwidth}
  \tablecolumns{9}
  \tablecaption{The computed combination and difference bands for naphthalene. Band positions determined using the 6-311G** basis set are compared to those determined using the 4-31G basis set. The forbidden- and allowed transitions that produce each band are given. Visualization of the fundamental vibrations producing the combination and difference bands indicated here in columns 4 and 7 are shown in Fig. \ref{fig:neutral} and \ref{fig:cation}. \label{tab:naphthalene}}
  \tablehead{
                             & \multicolumn{3}{c}{Forbidden transition} & & \multicolumn{4}{c}{Allowed transition}                             \\
  Wavelength                 & Wavelength       & Symmetry  & Fig.      & & Wavelength & Symmetry & Fig. & Intensity                           \\
  {[$\mu$m]/[cm$^{\rm -1}$]} & [$\mu$m]         &           &           & & [$\mu$m]   &          &      & [$\mathrm{Km}\ \mathrm{mole}^{-1}$] \\
  }
  \startdata
  \cutinhead{naphthalene 6-311G$^{\mathrm{**}}$ neutral}
  Combination & & & & & & & & \\
  5.989 / 1670 & 11.330 & B$_{\rm 2g}$ & b & \vline &        &             &   &       \\
  5.770 / 1733 & 10.570 & B$_{\rm 1g}$ & c & \vline &        &             &   &       \\
  5.660 / 1767 & 10.194 & B$_{\rm 2g}$ & d & \vline & 12.703 & B$_{\rm 3u}$ & a & 114.1 \\
  5.496 / 1820 & 9.680  & A$_{\rm g}$  & e & \vline &        &             &   &       \\
  5.104 / 1959 & 8.537  & A$_{\rm g}$  & f & \vline &        &             &   &       \\
  Difference & & & & & & & & \\
  5.977 / 1673 & 7.324  &  A$_{\rm g}$  & h & \vline &        &             &  &        \\
  5.583 / 1791 & 8.017  &  B$_{\rm 3g}$ & i & \vline &  3.291 & B$_{\rm 1u}$ & g & 64.1  \\ 
  5.356 / 1867 & 8.537  &  A$_{\rm g}$  & f & \vline &        &             &   &       \\   
  5.309 / 1884 & 8.660  &  B$_{\rm 3g}$ & j & \vline &        &             &   &       \\    
  \noalign{\vskip 10pt}
  5.894 / 1679 & 7.324  &  A$_{\rm g}$  &h & \vline &         &             &   &       \\ 
  5.510 / 1815 & 8.017  &  B$_{\rm 3g}$ &i & \vline & 3.266 & B$_{\rm 2u}$ & k & 49.1    \\ 
  5.289 / 1891 & 8.537  &  A$_{\rm g}$  &f & \vline &         &             &   &       \\ 
  5.243 / 1907 & 8.660  &  B$_{\rm 3g}$ &j & \vline &         &             &   &       \\ 
  \cutinhead{naphthalene 6-311G$^{\mathrm{**}}$ cation}
  Combination & & & & & & & & \\
  5.948 / 1681 & 21.365 & B$_{\rm 3g}$ & D & &  8.239  & B$_{\rm 2u}$ & A & 218.8  \\ 
  5.896 / 1696 & 10.751 & B$_{\rm 2g}$ & E & & 13.047  & B$_{\rm 3u}$ & B & 105.7  \\ 
  5.804 / 1723 & 19.626 & A$_{\rm g}$  & F & &  8.239  & B$_{\rm 2u}$ & A & 218.8  \\ 
  5.779 / 1730 & 10.366 & B$_{\rm 1g}$ & G & & 13.047  & B$_{\rm 3u}$ & B & 105.7  \\ 
  5.610 / 1783 &  9.826 & B$_{\rm 2g}$ & H & & 13.047  & B$_{\rm 3u}$ & B & 105.7  \\ 
  5.511 / 1815 &  9.541 & A$_{\rm g}$  & I & & 13.047  & B$_{\rm 3u}$ & B & 105.7  \\ 
  5.129 / 1950 & 23.198 & B$_{\rm 2g}$ & J & &  6.569  & B$_{\rm 1u}$ & C &  87.41 \\ 
  5.116 / 1955 & 13.499 & B$_{\rm 1g}$ & K & &  8.239  & B$_{\rm 2u}$ & A & 218.8  \\ 
  5.110 / 1957 & 8.410  & A$_{\rm g}$  & L & & 13.047  & B$_{\rm 3u}$ & B & 105.7  \\ 
  5.063 / 1975 & 13.131 & A$_{\rm g}$  & M & &  8.239  & B$_{\rm 2u}$ & A & 218.8  \\ 
  5.027 / 1989 & 21.365 & B$_{\rm 3g}$ & D & &  6.569  & B$_{\rm 1u}$ & C & 87.4   \\ 
  Difference & & & & & & & & \\
  5.415 / 1847 & 3.267  & A$_{\rm g}$  & N & \vline &       &             &   &      \\
  5.410 / 1848 & 3.266  & B$_{\rm 3g}$ & O & \vline & 8.239 & B$_{\rm 2u}$ & A & 218.8 \\
  5.371 / 1862 & 3.252  & B$_{\rm 3g}$ & P & \vline &       &             &   &       \\
  \noalign{\vskip 14pt}
  \cutinhead{naphthalene 4-31G neutral}\\
  Combination & & & & & & & & \\
  5.977 / 1673 & 11.296 & B$_{\rm 2g}$ & \nodata & \vline &        &             &         &      \\ 
  5.776 / 1731 & 10.600 & B$_{\rm 1g}$ & \nodata & \vline &        &             &         &      \\     
  5.637 / 1774 & 10.134 & B$_{\rm 2g}$ & \nodata & \vline & 12.677 & B$_{\rm 3u}$ & \nodata & 111.2 \\        
  5.521 / 1811 & 9.768  & A$_{\rm g}$  & \nodata & \vline &        &             &         &       \\  
  5.078 / 1969 & 8.479  & A$_{\rm g}$  & \nodata & \vline &        &             &         &       \\  
  Difference & & & & & & & & \\
  5.796 / 1725 &     7.386& A$_{\rm g}$  & \nodata & \vline  &       &             &         &      \\
  5.484 / 1823 &     7.965& B$_{\rm 3g}$ & \nodata & \vline  & 3.248 & B$_{\rm 1u}$ & \nodata & 78.1 \\
  5.264 / 1900 &     8.479& A$_{\rm g}$  & \nodata & \vline  &       &             &         &      \\
  5.212 / 1919 &     8.618& B$_{\rm 3g}$ & \nodata & \vline  &       &             &         &      \\ 
  \noalign{\vskip 10pt}
  5.738 / 1743 &    7.386 & A$_{\rm g}$  & \nodata & \vline  &       &             &         &       \\
  5.431 / 1841 &    7.965 & B$_{\rm 3g}$ & \nodata & \vline  & 3.229 & B$_{\rm 2u}$ & \nodata & 71.4  \\
  5.216 / 1917 &    8.479 & A$_{\rm g}$  & \nodata & \vline  &       &             &         &       \\
  5.165 / 1936 &    8.618 & B$_{\rm 3g}$ & \nodata & \vline  &       &             &         &       \\
  \enddata
\end{deluxetable}

\begin{deluxetable}{c@{\hspace{5mm}}ccccccc}
  \tablewidth{\textwidth}
  \tablecolumns{8}
  \tablecaption{The computed overtone, combination and difference bands for neutral coronene computed using the 4-31G basis set. The forbidden- and allowed transitions producing each band are given. \label{tab:coronene}}
  \tablehead{
                             & \multicolumn{2}{c}{Forbidden transition} & & \multicolumn{3}{c}{Allowed transition}                      \\
  Wavelength                 & Wavelength       & Symmetry              & & Wavelength & Symmetry  & Intensity                          \\
  {[$\mu$m]/[cm$^{\rm -1}$]} & [$\mu$m]         &                       & & [$\mu$m]   &           & [$\mathrm{Km}\ \mathrm{mole}^{-1}$] \\              
  }
  \startdata
  Overtone & & & & & & \\
  5.770 / 1733 & & & & 11.538& B$_{\rm 3u}$& 175.6 \\
  Combination & & & & & & & \\
  5.857 / 1707 & 11.890 & B$_{\rm 1g}$  & \vline &        &             &      \\
  5.723 / 1747 & 11.349 & B$_{\rm 2g}$  & \vline &        &             &      \\
  5.640 / 1773 & 11.041 & B$_{\rm 2g}$  & \vline &        &             &      \\
  5.475 / 1826 & 10.409 & B$_{\rm 1g}$  & \vline & 11.538 & B$_{\rm 3u}$ & 175.6 \\ 
  5.470 / 1828 & 10.394 & B$_{\rm 2g}$  & \vline &        &             &      \\
  5.432 / 1841 & 10.252 & B$_{\rm 2g}$  & \vline &        &             &      \\ 
  5.383 / 1858 & 10.091 & A$_{\rm g}$	& \vline &        &             &      \\
  5.281 / 1894 &  9.739 & A$_{\rm g}$   & \vline &        &             &      \\
  Difference & & & & & & \\
  5.917 / 1690 &     7.206 &  B$_{\rm 3g}$ & \vline &       &            &      \\
  5.916 / 1690 &     7.207 &  A$_{\rm g}$  & \vline &       &            &      \\
  5.732 / 1745 &     7.501 &  A$_{\rm g}$  & \vline &       &            &      \\
  5.460 / 1832 &     8.024 &  B$_{\rm 3g}$ & \vline &       &            &      \\
  5.421 / 1845 &     8.110 &  A$_{\rm g}$  & \vline & 3.249 & B$_{\rm 1u}$ & 140.3 \\
  5.399 / 1852 &     8.159 &  B$_{\rm 3g}$ & \vline &       &            &      \\
  5.396 / 1853 &     8.166 &  A$_{\rm g}$  & \vline &       &            &      \\
  5.253 / 1904 &     8.517 &  A$_{\rm g}$  & \vline &       &            &      \\
  5.252 / 1904 &     8.519 &  B$_{\rm 3g}$ & \vline &       &            &      \\
  \noalign{\vskip 10pt} 
  5.989 / 1670 &     7.024 &  B$_{\rm 3g}$ & \vline &       &            &       \\
  5.987 / 1670 &     7.026 &  A$_{\rm g}$  & \vline &       &            &       \\
  5.863 / 1706 &     7.206 &  B$_{\rm 3g}$ & \vline &       &            &       \\
  5.862 / 1706 &     7.207 &  A$_{\rm g}$  & \vline &       &            &       \\
  5.681 / 1760 &     7.501 &  A$_{\rm g}$  & \vline &       &            &       \\
  5.414 / 1847 &     8.024 &  B$_{\rm 3g}$ & \vline & 3.233 & B$_{\rm 2u}$ & 139.4 \\
  5.375 / 1860 &     8.110 &  A$_{\rm g}$  & \vline &       &            &       \\
  5.354 / 1868 &     8.159 &  B$_{\rm 3g}$ & \vline &       &            &       \\
  5.351 / 1869 &     8.166 &  A$_{\rm g}$  & \vline &       &            &       \\
  5.210 / 1919 &     8.517 &  A$_{\rm g}$  & \vline &       &            &       \\
  5.209 / 1920 &     8.519 &  B$_{\rm 3g}$ & \vline &       &            &       \\
  \enddata
\end{deluxetable}
  
\begin{figure}
  \plotone{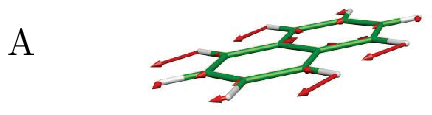}
  \caption{Example of on fundamental CH vibrational mode illustrative of the combination and difference bands of the naphthalene cation listed in \ref{tab:naphthalene}. The carbon skeleton is depicted in green, the hydrogen atoms in white and the vibrational motion of the molecule is indicated by the red vector arrows. Visualizations for all modes are available in the electronic edition of the Journal. \label{fig:cation}}
\end{figure}

The results for the 4-31G basis set are rather similar to those of the
larger basis set and are presented in the lower section of Table
\ref{tab:naphthalene}. The combination bands are in somewhat better
agreement than the difference bands, but overall the small basis set
results are in good agreement with those from the larger basis
set. Therefore we also computed the overtone, combination and
difference bands in neutral coronene with the 4-31G basis set and
those results are given in Table \ref{tab:coronene}. While coronene
has D6h symmetry, we use D2h symmetry labels to facilitate comparison
with naphthalene. In this way the similarities between the vibrational
symmetries are more readily apparent. It is important to note that
coronene has duo hydrogens, not quartet hydrogens as in
naphthalene. The CH$_{\rm oop}$ band in coronene falls at 11.538
$\mu$m (867 cm$^{\rm -1}$) whereas in naphthalene it is at 12.677
$\mu$m (789 cm$^{\rm -1}$). Consequently, coronene has an overtone
band in the 5 to 6 $\mu$m region derived from this strong
fundamental. Extrapolating to other PAHs, on the basis of typical band
positions for CH$_{\rm oop}$ bends, it is clear that PAHs containing
solo and duo hydrogens will have overtone bands in the 5 to 6 $\mu$m
region, a result consistent with the experimental conclusions in
Sect. \ref{sec:exp}. The second point to note from the compute
coronene tabulation is that all of the combination bands involve
CH$_{\rm ip}$ and CH$_{\rm oop}$ bending modes coupled to the one
strong out-of-plane mode. As for naphthalene, the difference bands
arise from a coupling of a strong CH stretch with forbidden CH$_{\rm
  ip}$ bending modes.

The naphthalene cation was also studied using the {6-311G**} basis set
and the results are given in the middle section of Table
\ref{tab:naphthalene}, with visualizations of the modes in
Fig. \ref{fig:cation}. In contrast with neutral naphthalene, the first
point to note is that, since ionization increases the intrinsic
strength of many bands, the cation has several strong bands that can
contribute to combination bands. For naphthalene, the contribution
bands are mostly CH$_{\rm ip}$ and CH$_{\rm oop}$ bending in
character, but some begin to have contributions from modes with CC
stretching and bending character. For example, the bands at 23 and 21
$\mu$m (435 and 476 cm$^{\rm -1}$) have significant CC contributions
to the mode. As the cation grows in size and the number of low
frequency modes increases, we presume that the character of the
vibrations that produce combination bands in the cation could become
even more diverse. The second difference between the neutral and
cation forms is that the cation difference bands involve an allowed CH
bend and forbidden CH stretches. That is, the same strong CH$_{\rm
  ip}$ bend that contributes to the combination bands is also the
allowed band contributing to the difference bands.

Comparing the position of the computed combination bands with the
frequencies of the two fundamentals that comprise the bands shows only
a small shift (2.6 cm$^{\rm -1}$) for the region of interest. This,
coupled with the spectroscopic property that difference bands fall at
($\nu_{i}- \nu_{j}$) without any shift \citep{1945.book.....H},
permits using scaled harmonic frequencies from larger PAHs as a guide
in assessing the fundamental bands that can contribute to the observed
overtone, difference, and combination bands. Considering the modes in
three very large regular PAHs, C$_{\rm 110}$H$_{\rm 26}$, C$_{\rm
  112}$H$_{\rm 26}$, and C$_{\rm 130}$H$_{\rm 28}$, we find that the
cation, neutral, and anion forms of all three molecules should each
have strong overtones near 5.4 - 5.6 $\mu$m (1818 cm$^{\rm
  -1}$). Considering that we are computing the overtone as twice the
scaled harmonics, the 5.4 - 5.6 $\mu$m bands are consistent with the
astronomical 5.7 $\mu$m feature as arising from overtones of large
symmetric PAHs. All three charge states of all three molecules have
many possible combination and difference bands in the 5 - 6 $\mu$m
region. For the neutral combination bands, the strongest allowed
component are CH$_{\rm oop}$ bends, the same as found for the smaller
molecules. However, it is interesting to note that there are
combination bands in this region that are derived from an allowed
CH$_{\rm ip}$ bend with a forbidden CH$_{\rm oop}$ bend and an allowed
CC-stretch with a forbidden drum head mode; the intensity of the
allowed CH$_{\rm ip}$ bend and CC stretch is about 25\% of the
CH$_{\rm oop}$ bend. For anions and cations, the intensity of the
CH$_{\rm ip}$ and CC stretch can exceed the strongest CH$_{\rm oop}$
bending vibration. For large PAH cations, the CH-stretch intensity can
be sizable, therefore difference bands in the 5 - 6 $\mu$m region can
arise from an allowed CH stretch as well as from an allowed CH
bend. Overall, the anion spectra in the 5 - 6 $\mu$m region is more
similar to that of the cation than that of the neutral.

Summing up, the computational work supports the idea that PAH
neutrals, cations, and anions have overtones arising from CH$_{\rm
  oop}$ bends of solo and duo hydrogens that fall in the 5 - 6 $\mu$m,
as do combination bands arising from in- and out-of-plane CH bends. In
addition to the combination bands, difference bands, involving CH
stretches and CH bends, can also fall in this region. Since there are
more strong bands in the cations and anions than in the neutrals, more
combination bands arise. It is likely that in large PAH ions, not all
of the combination bands arise from only CH$_{\rm ip}$ and CC$_{\rm
  oop}$ bends.

For now our theoretical study is limited by the inability to compute
intensities. Without intensities the theory is no up to making a
one-to-one comparison with the laboratory spectra and it is impossible
to definitively establish which combination and difference bands could
contribute to the astronomical spectrum. However, the calculations
support the idea that neutrals, cations, and anions of large PAHs can
all contribute to the observed features. Specifically: 1) the CH$_{\rm
  oop}$ solo and duo hydrogens lead to an overtone in the region of
interest, 2) combination bands arising from in- and out-of-plane CH
bends can also contribute to the observed bands, and 3) difference
bands involving CH stretches and CH bends can also contribute to the
observed bands.

\section{Astrophysical  Considerations}
\label{sec:astrophysical}

\begin{figure}
  \plotone{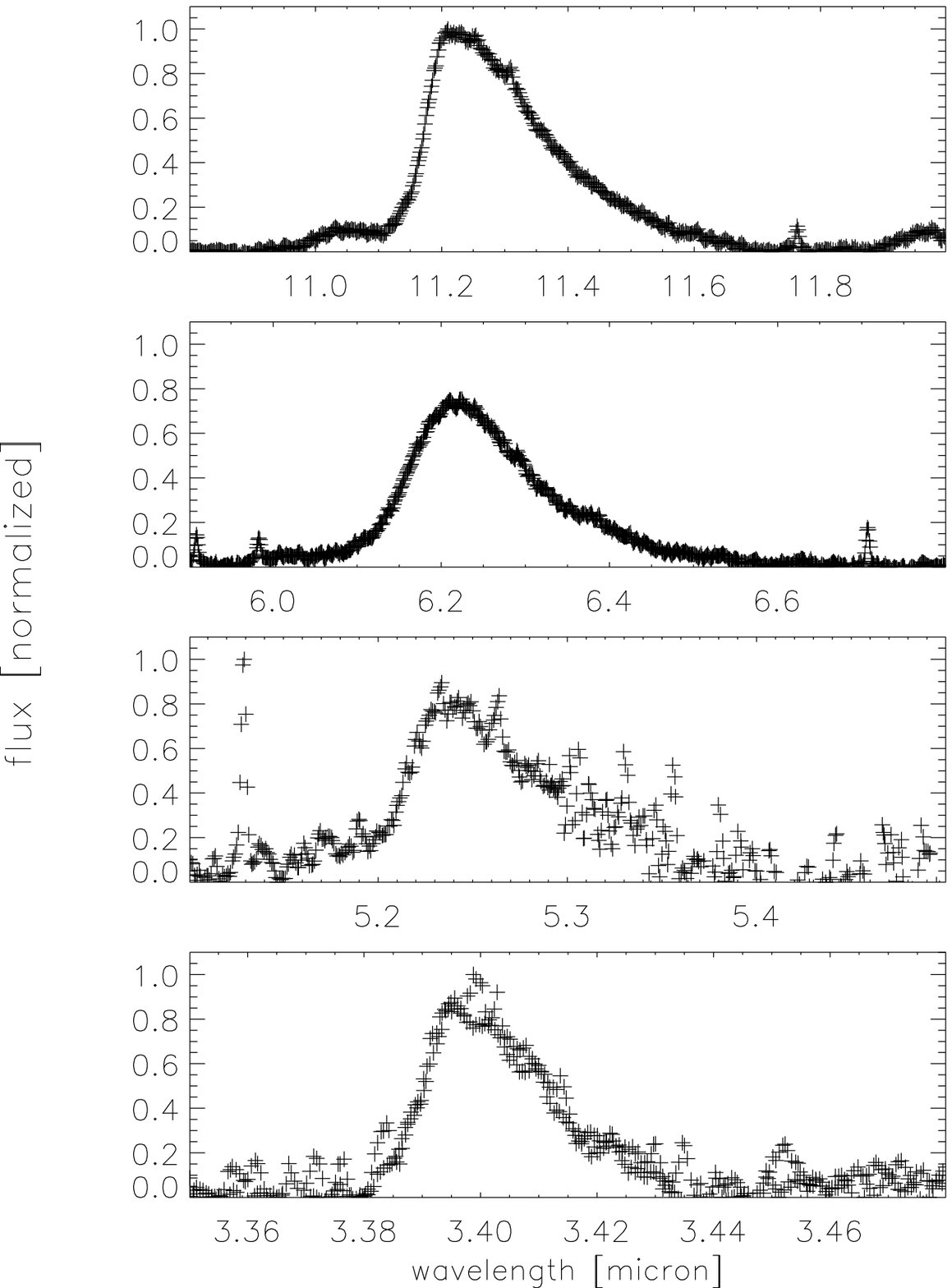}
  \caption{Comparison between the normalized profile of the 5.25 $\mu$m band to those of the normalized 3.4, 6.2 and 11.2 $\mu$m bands in NGC 7027, note the similarities.\label{fig:profiles}}
\end{figure}

First we consider the 5.25 $\mu$m band. The few earlier studies of
this band focused on its general
attributes. \cite{1989ApJS...71..733A} and \cite{1996MNRAS.281L..25R}
attributed this feature to overtones and combinations involving the
lower lying CH$_{\rm oop}$ bending modes. For three of the four
objects considered here the 5.25 $\mu$m emission band has a FWHM of
0.11 $\mu$m (40 cm$^{\rm -1}$). It may be slightly broader in NGC 7027
(0.14 $\mu$m, 50 cm$^{\rm -1}$), but the lower signal to noise in this
object makes this value less reliable. \cite{1996MNRAS.281L..25R}
noted the striking similarity between the profiles of the 5.25 $\mu$m
band and that of the 3.4, and 11.2 $\mu$m bands. This comparison is
revisited in Fig. \ref{fig:profiles}, comparing the 5.25 $\mu$m band
with the 3.4, 6.2, and 11.2 $\mu$m emission bands from NGC 7027 using
the higher quality ISO SWS data. The major astronomical features at
6.2 and 11.2 $\mu$m arise from fundamental vibrations whereas the 5.25
$\mu$m band is produced by combinations and overtones involving
fundamentals. The similarities between the profiles of these bands are
striking, even down to the blue satellite features for the 6.2 and
11.2 $\mu$m bands. However, on an (absolute) energy scale the widths
and skewness of the profiles show some variations, with the 6.2 $\mu$m
profile clearly different. Both width and skewness are tied to the
anharmonicity parameter of the transition involved, which depends on
the involved mode \citep{1995A&A...299..835J}. While presently
relevant spectroscopic data is lacking, we note that future
experimental studies on the relevant modes may prove to be very
insightful.

\begin{figure}
  \plotone{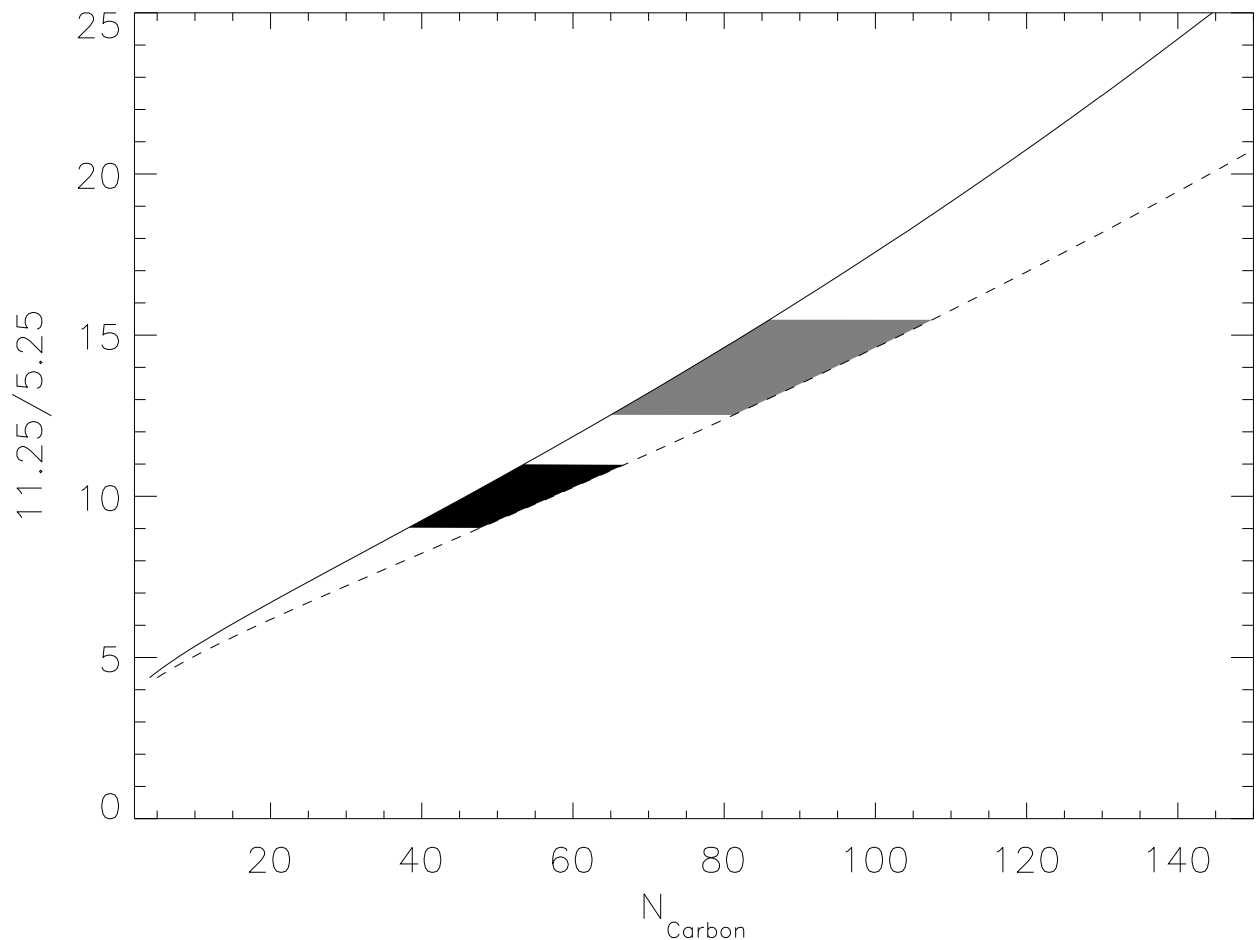}
  \caption{The I$_{11.25}$/I$_{5.25}$ ratio as a function of neutral PAH size modelled using the thermal approximation and including radiative cooling. The solid line is the ratio for an average photo-excitation energy of 8 eV and the dashed line for an average photon energy of 10 eV. The black area is where the astronomically observed ratio of $\sim$10 ($\pm$1.0) falls and the area in gray that where the observed ratio of $\sim$14 ($\pm$1.5).\label{fig:size}}
\end{figure}

\cite{1996MNRAS.281L..25R} found that the ratio of the 11.2 $\mu$m
band intensity to that of the 5.25 $\mu$m band was roughly constant
(28 - 32) over their sample of three PNe. Here, from Table
\ref{tab:strengths} the I$_{11.25}$/I$_{5.25}$ in HD 44179 and the two
positions in Orion are about 10 ($\pm$ 1.0), whereas it is 14 ($\pm
1.5$) in NGC 7027. We have developed a simple model to computationally
determine this ratio based on the average ($\sim$ 9.5), experimentally
established, (10 - 15) / 5.25 $\mu$m cross sections of 15 neutral PAHs
with 28 - 50 carbon atoms. The model predicts the
I$_{11.25}$/I$_{5.25}$ ratio as a function of PAH size and photon
excitation energies. The thermal approximation is used to determine
the emission spectrum and radiative relaxation is taken into account
\citep{1997ApJ...475..565D, 2001ApJ...556..501B}. Figure
\ref{fig:size} shows the predicted ratios as a function of PAH size
between two different excitation energies, 8 and 10 eV. For average
photon excitation energies of 8 and 10 eV, PAHs with about 40 to 65
carbon atoms produce the observed ratios for HD 44179 and the two
positions in Orion. NGC 7027 implies that PAHs with between 65 to
about 105 carbon atoms are required to match the observed ratio. This
is very much in line with the estimate based on the 3.3/11.2 $\mu$m
ratio \citep{1989ApJS...71..733A}.

\begin{figure*}
  \plottwo{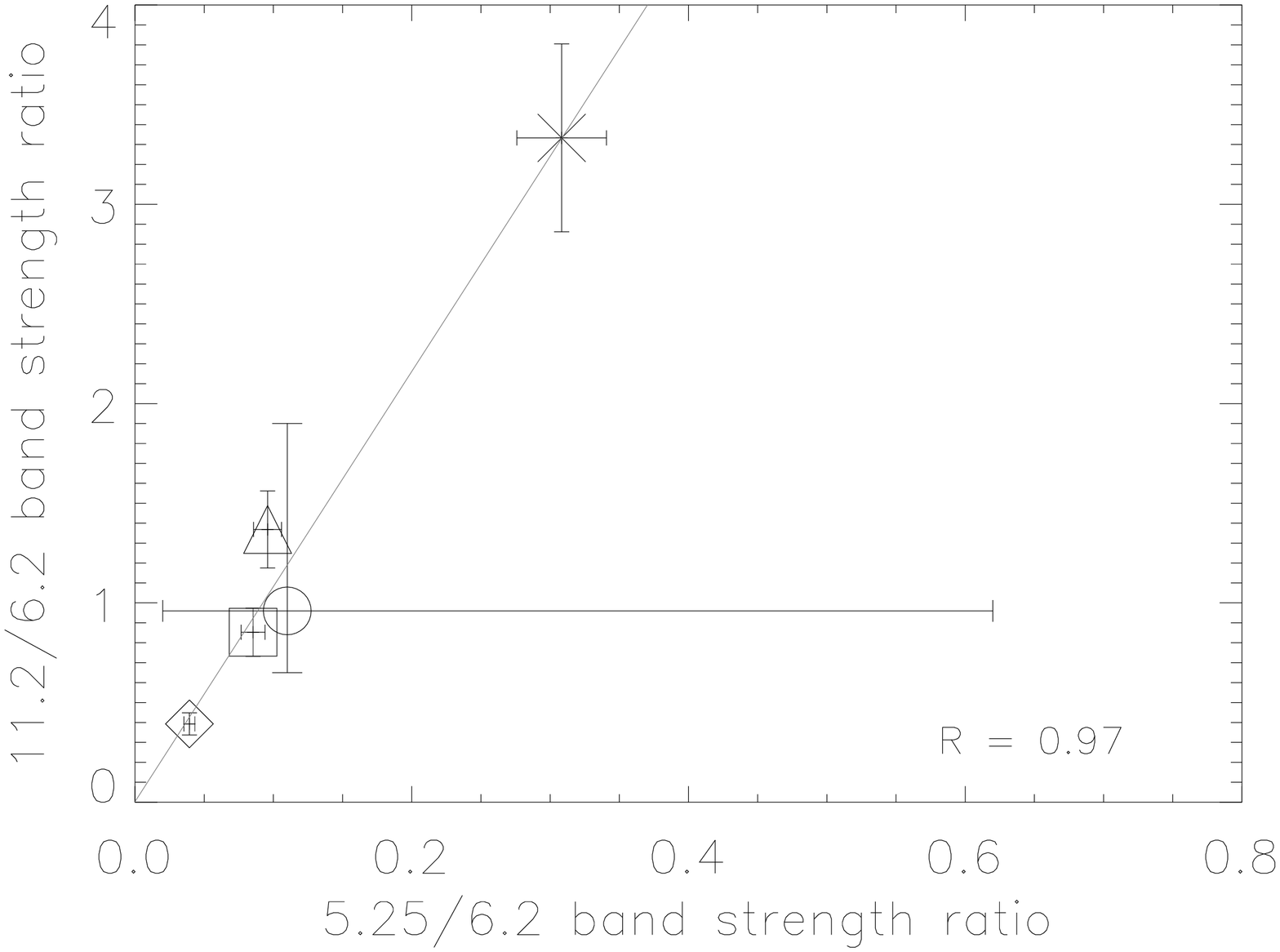}{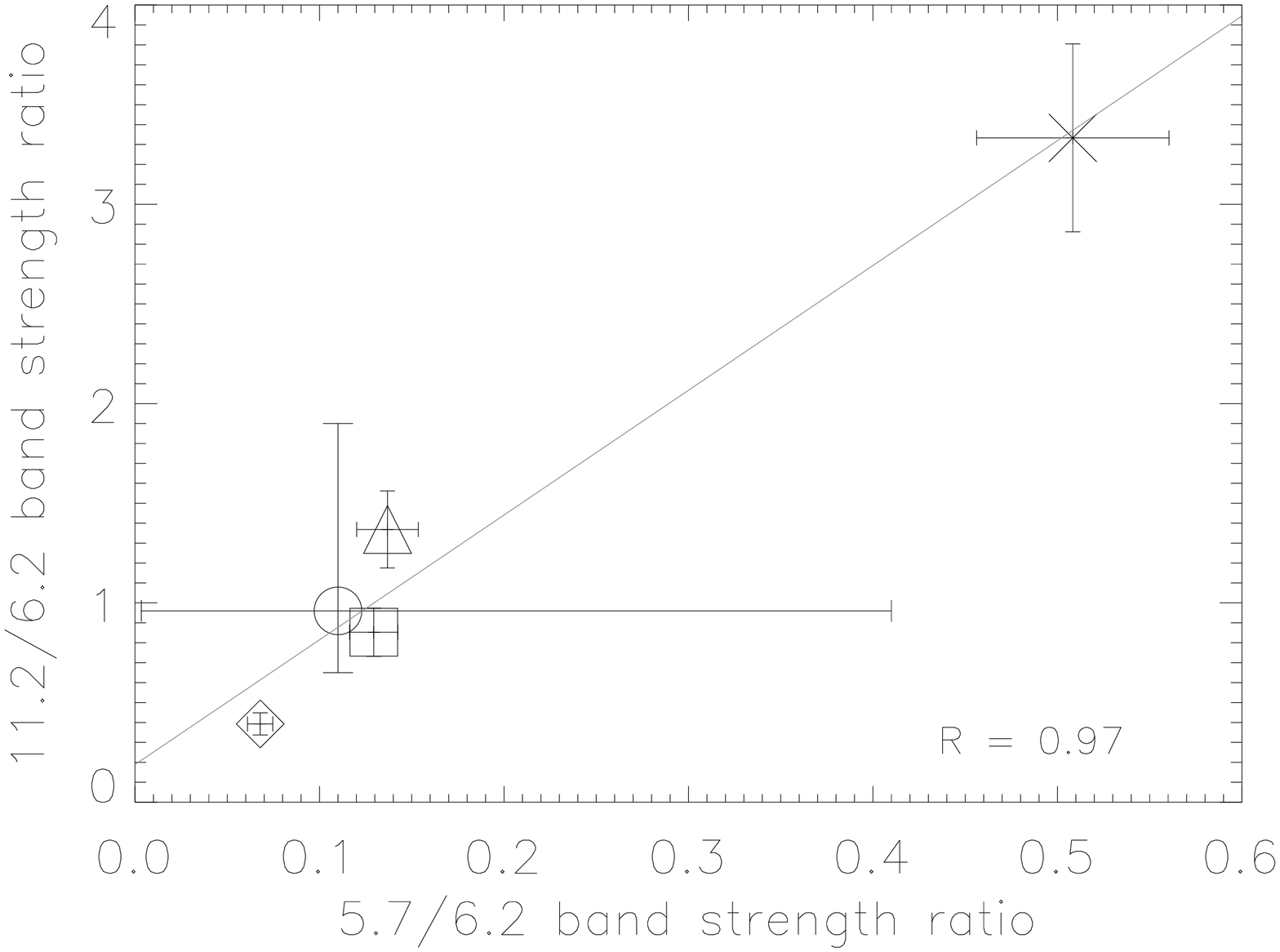}
  \caption{Band strength ratio correlations. $\diamond$ HD 44179; $\vartriangle$ NGC 7027; $\square$ Orion Bar H2S1; $\times$ Orion Bar D5; $\circ$ Nearby galaxies (Smith, J.~D., private communication). Note that for the Galaxies the strength of the 5.25 $\mu$m band was determined by assuming a slightly asymmetric Drude profile with a width and wavelength center trained on a high S/N subset of the SINGS \citep{2003PASP..115..928K}  spectra. Integrating the full profile produces the used bands strength. Also note that some of the SINGS spectra have modest redshifts, which bring slightly more of the feature in view, up to roughly a quarter of the feature's FWHM. \label{fig:ratios}}
\end{figure*}

The analysis of the laboratory data and the quantum-chemical
calculations point to the involvement of CH$_{\rm oop}$ modes for the
emission between 5 - 6 $\mu$m. The good correlation between the
5.25/6.2 with the 11.2/6.2 $\mu$m band strength ratios (R = 0.97) and
the 5.7/6.2 with the 11.2/6.2 $\mu$m band strength ratios (R = 0.97),
which traces the CH modes, affirm this (Fig. \ref{fig:ratios}). In
contrast, there is no good correlation between the 5.25/11.2 with the
6.2/11.2 $\mu$m ratios. Showing again that the 5.25 $\mu$m and CC
modes are not intimately connected in the astronomical spectra.

Ionization of PAH molecules can influence band positions, but not as
notably as the effect on the relative band intensities. Upon
ionization the strength of the CC modes increases significantly, while
the CH modes tend to decrease in strength. However, there appears to
be a size dependence to this. PAHs of a comparable to those implied
here do not have significant reduction in CH band strength
\citep{2008ApJ...678..316B}. NGC 7027 and the two positions in the
Orion Bar probe the two extremes in ionizing environments. Future
studies on a larger sample will have to establish whether the
difference in 5.25/11.2 $\mu$m ratio for NGC 7027 is a real property
of the interstellar PAH spectra. While the theoretical data suggest
that ionized PAHs and non-CH$_{\rm oop}$ modes may be involved, for
now we conclude that the correlation in Fig. \ref{fig:ratios}
advocates for neutral carriers dominating the astronomical 5.25 $\mu$m
PAH band.

\cite{2001A&A...370.1030H} made a study of the 10 - 15 $\mu$m region,
the bands which are due to aromatic CH$_{\rm oop}$ vibrational
modes. Laboratory data and quantum-chemical calculations show that the
peak position of the bands in this region depends on the number of
adjacent hydrogen atoms on the periphery of PAH molecules. Therefore,
ratios of these bands will reveal the structure of the emitting PAHs
in astronomical environments. The predominance of the 11.2 over 12.7
$\mu$m band intensity in the four sources investigated here points to
compact carriers \citep{2001A&A...370.1030H}. Since the emission in
the 5 - 6 $\mu$m region involves coupling with, and intensity
borrowing from the CH$_{\rm oop}$ modes, this also holds for the 5.25
and 5.7 $\mu$m bands. However, it is prudent to look at the 5 - 6
$\mu$m region of sources in which the 12.7 is as strong as the 11.2
$\mu$m band (e.g. IRAS 18317). Unfortunately good signal-to-noise
spectra of the 5.25 $\mu$m band are not yet available for the sources
known to show such ratios.

Lastly, the presence of quartets produces emission in-between and
beyond the positions of the observed astronomical 5.25 and 5.7 $\mu$m
features excludes them as important part of the astronomical PAH
family \citep{2001A&A...370.1030H}. This is completely consistent with
the conclusions about the PAH structures from the 10 - 15 $\mu$m
region.

We turn now to the 5.7 $\mu$m band. Figure \ref{fig:adjacency}
illustrates the dependence of the 5.7 $\mu$m peak position on the
molecular edge structure. As the laboratory data show, duos and trios
cluster on the blue side whereas the presence of solos produce
features on the red side of the feature. While the four spectra are
quite similar, the spectrum of the Orion Bar position D5 seems to
indicate that the duo and trio hydrogens are as important as solo
hydrogens, even though both Orion Bar positions show similar 11.2/12.7
$\mu$m band strength ratios. The difference in the spectra between
Orion Bar positions D5 and H2S1 suggest a different PAH population at
both positions. This difference might be driven by erosion, where more
irregular shaped PAHs are present at position D5.

All of the fifteen initially considered astronomical spectra show
emission features at 5.25 and 5.7 $\mu$m, but nothing significant
in-between. This rules out a large population of quartet containing
PAHs. The spectra span a significant range of astrophysical
environments, suggesting that many, if not most, of the astronomical
PAH population is dominated by large and compact PAHs. It bears
repeating here that the greater intrinsic strength of the solo CH over
that of duo CH is largely responsible for the predominance of the 11.2
$\mu$m astronomical band in the CH$_{\rm oop}$ region.

It has been shown by \cite{2002A&A...390.1089P} that the variations in
peak position and shape of the PAH bands in the 6 - 9 $\mu$m region
can be grouped into distinctive classes. Moreover, these classes seem
to correlate with object type. All ISM-like sources belong to class A.
These are characterized by a CC-stretch band peaking at 6.2 $\mu$m and
the `7.7' $\mu$m band complex peaking at 7.6 $\mu$m. On the other
hand, isolated Herbig Ae/Be stars, together with a few Post- AGB stars
and most PNe belong to class B. These are characterized by a
CC-stretch band peaking between 6.24 and 6.28 $\mu$m and `7.7' $\mu$m
band complex peaking between 7.8 and 8.0 $\mu$m. HD 44179, NG7027 and
the two positions in the Orion Bar cover both of these classes, with
HD 44179 and NGC 7027 representing class B and both positions in the
Orion Bar representing class A. Interestingly enough, for the quality
of the spectra available, the 5.25 and 5.7 $\mu$m features do not
allow for such a classification, behavior consistent with CH
modes. Given that the shift between class A and class B type of
spectra seems strongly influenced by the relative populations of PAH
cations to PAH anions \citep{2008ApJ...678..316B}, the apparent
independence of the astronomical 5.25 and 5.7 $\mu$m bands from these
classifications strongly suggests again that they track the neutral
PAH population.

The spectrum labeled `Galaxies' in Fig. \ref{fig:compare} presents the
slightly redshifted average galaxy spectrum from
\cite{2007ApJ...656..770S}, obtained using Spitzer's Infrared
Spectrograph \citep[IRS;][]{2004ApJS..154...18H} and shows part - due
to IRS' cutoff at 5.2 $\mu$m - of the 5.25 $\mu$m band and the 5.7
$\mu$m band. Compared to ISO, Spitzer has limited spectral resolution
but superior sensitivity. The spectrum demonstrates the accessibility
of this region using Spitzer and therefore the possibility for band
strength analysis and, for the 5.7 $\mu$m band, studies of erosion
processes as outlined in this paper.

\section{Summary and Conclusion}
\label{sec:summary}

This paper presents a study of the two minor PAH features in the 5 - 6
$\mu$m (2000 - 1667 cm$^{\rm -1}$) region, centered at positions near
5.25 (1905 cm$^{\rm -1}$) and 5.7 $\mu$m (1739 cm$^{\rm -1}$). These
contain information about the interstellar PAH population and
conditions in the emission regions that both complement and extend the
information revealed by the major bands.

Fifteen high quality ISO SWS spectra have been investigated for
emission in the this wavelength region, with four spectra having
sufficient signal-to-noise to allow for an in-depth analysis. Combined
with a spectral database comprised of laboratory studies and dedicated
quantum-chemical calculations, these spectra allow us to probe the
main characteristics of the carriers of the astronomical PAH
features. After continuum and emission line removal, all four
astronomical spectra show similar, almost universal profiles. However,
the signal-to-noise level can be improved and there are hints of
subtle, but interesting, variations.

The absence of bands between 5 - 6 $\mu$m in laboratory spectra of
deuterated polycyclic aromatic molecules, as well as the absence of
fundamentals in the quantum-chemical calculations in this region along
with the strong correlation between the 5.25 and 5.75 $\mu$m band
strength with the 11.2 $\mu$m band strength in the astronomical
spectra, substantiates the involvement of CH$_{\rm ip}$ and CH$_{\rm
  oop}$ bending vibrations. In-depth analysis of the laboratory
spectra and quantum-chemical calculations show that the astronomical
5.25 and 5.75 $\mu$m bands are a blend of combination, difference, and
overtone bands, involving CH$_{\rm ip}$ and CH$_{\rm oop}$ bending and
stretching modes and, likely for the larger ionized PAHs, CC$_{\rm
  oop}$ modes. When it becomes possible to compute the intensities of
overtone and combination bands, this work should be extended.

Turning to the hydrogen adjacency classes, PAHs with solo and duo
hydrogens consistently produce prominent bands in the appropriate
wavelength regions, whereas PAHs with higher adjacency hydrogens show
far richer spectra. These produce bands in-between and beyond the 5.25
and 5.7 $\mu$m bands, ruling such species out as important members of
the emitting population. The 5.7 $\mu$m feature in itself - through
its profile - contains adjacency class information that might be more
easily accessible than through the 10 - 15 $\mu$m CH$_{\rm oop}$
region, where it is difficult to separate the duo and trio hydrogen
modes.

The data suggest that the emitting astronomical PAHs are mostly large
(50 $\simeq N_{\rm C}\simeq$ 100 ), compact, and not fully
deuterated. Furthermore, both the quantum-chemical calculations and
the absence of a correlation between the 5.25/6.2 and 5.7/6.2 $\mu$m
band strengths with the 11.2/6.2 $\mu$m band strength ratio suggest
that the 5.25 and 5.7 $\mu$m PAH band do not trace ionization and are
carried predominately by neutrals. This point is reinforced by the
lack of connection with class A and class B PAH band behavior. This
suggests that high quality spectra from 5 - 10 $\mu$m provide insight
into the neutral as well as the cation and anion members of the
emitting astronomical PAH family. Even so, a bigger collection of
spectra of large compact PAHs, with mixtures of solo, duo and trio
hydrogens, is still needed to firm up these conclusions.

\acknowledgments

Christiaan Boersma acknowledges support from the Netherlands
Organization for Scientific Research (NWO; Grant R 78-405). The
experimental aspects of this work were supported through NASA's Long
Term Space Astrophysics (Grant \# 907524) and Astrobiology (Grant \#
811073) Programs and earlier support from NASA's Laboratory
Astrophysics Program. Andrew Mattioda acknowledges the support of the
National Research Council. As always, we are deeply indebted to Robert
Walker for his outstanding technical support of all phases of the
experimental work.

\clearpage

\bibliographystyle{apj}
\bibliography{aamnem99,bibliography}

\end{document}